%% file: main.tex
\documentclass[a4paper]{article}
\usepackage[left=3cm,right=3cm,top=2cm,bottom=2.5cm]{geometry}
\usepackage{times}
\usepackage{authblk}
\usepackage{graphicx}
\usepackage{amsmath, amssymb}
\usepackage{booktabs}
\usepackage{color}
\usepackage[usenames,dvipsnames]{xcolor}
\usepackage[round]{natbib}

\usepackage{fancyhdr,graphicx,amsmath,amssymb}
\usepackage[ruled,vlined]{algorithm2e}

\usepackage[]{graphicx}
\chardef\bslash=`\\

\numberwithin{equation}{section} 
\usepackage{hyperref} 

\begin{document}

\title{A Balanced Statistical Boosting Approach \\ for GAMLSS via New Step Lengths}
\date{}
\author[$1$,*]{Alexandra Daub}
\author[$2$]{Andreas Mayr}
\author[$1$]{Boyao Zhang}
\author[$1$]{Elisabeth Bergherr}
\affil[$1$]{ \small Chair of Spatial Data Science and Statistical Learning, Georg-August-Universit\"at G\"ottingen, \hspace*{1em} \newline
\hspace*{-17.6em} Platz der G\"ottinger Sieben 3, 37073 G\"ottingen, Germany} 
\affil[$2$]{ \small Department of Medical Biometrics, Informatics and Epidemiology, University Hospital Bonn, \hspace*{0.2em} \newline
\hspace*{-22em} Venusberg Campus 1, 53127 Bonn, Germany \hspace*{22em}}
\affil[ ]{\hspace*{-20em} $^*$ \small Correspondence: alexandra.daub@uni-goettingen.de \hspace*{20em}}
\maketitle  
\vspace*{-3em}

\begin{abstract}
\footnotesize{\input{abstract}}
\end{abstract}
\input{introduction}

\input{methods}
\input{simus}
\input{application}

\input{conclusion}
\subsection*{Funding} 
This work was supported by the DFG (Deutsche Forschungsgemeinschaft; Projekt BE 7939/2-2)
and the Volkswagen Foundation, project ``Bayesian Boosting - A new approach to data science, unifying two statistical philosophies''. 
\bibliographystyle{apalike} 
\bibliography{literature_paper1}

\newpage

\appendix
\input{appendix}

\end{document}

%% file: abstract.tex
Component-wise gradient boosting algorithms are popular for their intrinsic variable selection and implicit regularization, which can be especially beneficial for very flexible model classes.
When estimating generalized additive models for location, scale and shape (GAMLSS) by means of a component-wise gradient boosting algorithm, an important part of the estimation procedure is to determine the relative complexity of the submodels corresponding to the different distribution parameters. 
Existing methods either suffer from a computationally expensive tuning procedure or can be biased by structural differences in the negative gradients' sizes, which, if encountered, lead to imbalances between the different submodels. 
Shrunk optimal step lengths have been suggested to replace the typical small fixed step lengths for a non-cyclical boosting algorithm limited to a Gaussian response variable in order to address this issue.
In this article, we propose a new adaptive step length approach that accounts for the relative size of the fitted base-learners to ensure a natural balance between the different submodels.
The new balanced boosting approach thus represents a computationally efficient and easily generalizable alternative to shrunk optimal step lengths.
We implemented the balanced non-cyclical boosting algorithm for a Gaussian, a negative binomial as well as a Weibull distributed response variable and demonstrate the competitive performance of the new adaptive step length approach by means of a simulation study, in the analysis of count data modeling the number of doctor's visits as well as for survival data in an oncological trial. \\[-0.5em]

\noindent \textit{\textbf{Keywords}}: gradient boosting, GAMLSS, step length, variable selection, high-dimensional data  

%% file: introduction.tex
\section{Introduction}

Generalized additive models for location, scale and shape \citep[GAMLSS,][]{Rigby2005} are a flexible tool to model distribution parameters beyond the mean and as such extend the class of generalized additive models \citep[GAM,][]{Hastie1990}.
For example for modeling weather extremes \citep{Villarini2011,Lopez2013}, infant growth \citep{Villar2014, Cole2013} or key pandemic indicators \citep{Karim2021, Fritz2021}, 
GAMLSS allow to explain not only the mean but also the variation as well as other parameters of the response variable by means of covariates and therefore gained a lot of attention in statistical modeling \citep{Kneib2013, Klein2024}.
In general, GAMLSS consist of one additive submodel for each distribution parameter, where all submodels combined result in a response variable distribution whose form depends on the covariates. \\
In order to estimate GAMLSS, \cite{Rigby2005} proposed a penalized maximum likelihood approach.
Other estimation methods are based on Bayesian inference \citep{Klein2015, Umlauf2018} or statistical boosting; \cite{Mayr2012} for example introduced a component-wise gradient boosting algorithm for GAMLSS.
Component-wise gradient boosting algorithms \citep{Buehlmann2003, Buehlmann2007} update the model by a small portion of the best-performing so-called base-learner, e.g. fitted simple linear models or splines with a low degree of freedom.
The update procedure is performed iteratively until a predefined stopping iteration is reached, which constitutes the main tuning parameter.
The resulting model is hence an ensemble of weak learners \citep{Mayr2014}. 
In addition to a natural coefficient shrinkage and a higher prediction accuracy compared to unregularized estimation methods, component-wise gradient-boosting approaches offer an intrinsic variable selection as well as the opportunity to handle high-dimensional data \citep{Buehlmann2006}. 
Due to the larger overall complexity of GAMLSS compared to a classical GAM, the intrinsic variable selection and regularization provided by boosting approaches can be particularly useful for this model class. \\ 

When boosting GAMLSS, different submodels have to be updated within the iterative boosting routine, which can be done in different ways.
\cite{Mayr2012} proposed to update the different predictors one by one, cycling through the distribution parameters, i.e. every predictor is updated in every iteration.
The main additional task when boosting GAMLSS compared to models with a single predictor is to determine the relative complexity of the different submodels, which \cite{Mayr2012} addressed by determining a different stopping iteration for each submodel in a computationally demanding tuning procedure. 
In an alternative, non-cyclical, boosting algorithm that only updates one predictor in every iteration, \cite{Thomas2018} internalized the procedure of determining the relative importance of the different submodels and thus reduced the tuning procedure to a one-dimensional optimization problem.
As \cite{Zhang2022} showed exemplarily for a Gaussian response variable with a high variance, the selection of the submodel to update in Thomas et al.'s (2018) boosting algorithm can however be deterred by structurally differently sized negative gradients, 
which can moreover result in very long run times.
For example when modeling the malnutrition of children in India \citep{Fahrmeir2011}
based on the the mother's and the child's BMI as well as age, the non-cyclical boosting algorithm selects all four covariates for $\sigma$ while only the child's age is included in $\mu$ \citep{Zhang2022}.
Similar problems can arise when using the non-cyclical boosting algorithm in related model classes like distributional copula regression \citep{Hans2022} or multivariate GAMLSS \citep{Stroemer2023}. 
In order to address both, the imbalancedness issue as well as the problem of very long run times in the special case of a Gaussian location and scale model, \cite{Zhang2022} proposed to use shrunk optimal step lengths instead of small fixed step lengths for obtaining the (potential) predictor updates, 
where the optimal step lengths are either determined by line search or based on analytically derived solutions. \\

In this work, we propose a new adaptive step length approach for ensuring a natural balance between the different submodels. 
That allows on the one hand for easier generalization to other response variable distributions or model classes than analytically derived optimal step lengths do. 
On the other, it is computationally more efficient than using numeric optimization techniques. 
Concretely, we construct adaptive step lengths based on the relative size of different fitted base-learners in a way that effectively fixes all competing potential updates of the different submodels to the same size. 
In addition to equipping a fixed step length approach with this feature, we also combine shrunk optimal step lengths with the new base-learner based approach. 
We therefore derive the approximate analytical optimal step length for one distribution parameter for each non-Gaussian response variable distribution we consider. 
In addition to the case of a continuous Gaussian response variable, a negative binomial location and scale model for count data and a Weibull scale and shape model for survival data is considered. \\ 

The remainder of this article is structured as follows. 
In section~\ref{daub:section_methods}, we first introduce GAMLSS as well as component-wise gradient boosting approaches and give an overview of the existing algorithms for boosting GAMLSS. 
Then, the idea of shrunk optimal step lengths for non-cyclical boosting algorithms is discussed and the new adaptive step length approach including its motivation is presented. In addition to that, we derive one distribution parameter's approximate optimal step length in the negative binomial location and scale as well as in the Weibull model scale and shape model, respectively.
In section~\ref{daub:section_simulations}, results of a simulation study that comprises settings with structural differences in the negative gradients' scales for a Gaussian as well as a negative binomial response variable are presented.  
Real-world applications are considered in section~\ref{daub:section_applications} and section~\ref{daub:section_conclusion}~concludes with a summary of the main findings and a discussion.

%% file: methods.tex
\section{Methods}
\label{daub:section_methods}

\subsection{Boosting GAMLSS}
In GAMLSS, each distribution parameter $\theta_k, k \in \{1, ..., K\}$, is linked to the predictor $\eta_{\theta_k}$ via a known monotonic link function $g_k(\cdot)$ and the predictors are additively combined out of the effects of different covariates.
GAMLSS thus model the response variable $y$ by $y \, \vert \, \boldsymbol{x} \sim \mathcal{D}(\theta_1, ..., \theta_K)$ with 
\begin{align*}
g_k(\theta_k) = \eta_{\theta_k} = \beta_{0,\theta_k} + \sum_{j=1}^{J_k} f_{j,\theta_k}(x_{kj}) \, , \,\, k \in \{1, ..., K\} \, ,
\end{align*} 
where $x_{k1}, ..., x_{kJ_k}$ are the $J_k$ covariates corresponding to $\theta_k$, $\beta_{0,\theta_k}$ refers to the intercept of the $k$th submodel and the function $f_{j,\theta_k}(\cdot)$ represents the assumed type of effect between covariate $j$ and predictor $\eta_{\theta_k}$. 
In general, different covariate effects, for example linear, smooth, spatial or random effects can be considered. 
Within this work, we will focus on linear effects and will therefore have $f_{j,\theta_k}(x_{kj}) = x_{kj} \beta_{kj}$ 
for all $j \in \{1, ..., J_k\}$ and $k \in \{1, ..., K\}$. 
Each covariate $x_{kj}$ is observed for $n$ individuals resulting in an $n$-dimensional vector of realizations of every predictor $\eta_{\theta_k}$ and of each distribution parameter $\theta_k, k \in \{1, ..., K\}$, when fitting the model. \\

In order to incorporate an intrinsic variable selection and to handle high-dimensional data, \cite{Mayr2012} introduced a component-wise gradient boosting algorithm for GAMLSS as an alternative estimation method. 
In component-wise gradient boosting \citep{Buehlmann2003}, so called base-learners, e.g., linear models or smoothing splines with a low degree of freedom, are iteratively fitted to the negative gradient vector of the loss function. 
The best-performing base-learner is then scaled by a small step length factor and added to the model in every iteration.
As the same base-learner can be selected multiple times, the covariate effects grow over the iterations.
In order to prevent overfitting and improve the prediction accuracy, the iterative updates are stopped before convergence, which yields a natural shrinkage in the estimated covariate effects \citep{Mayr2014}.  
As opposed to general statistical boosting algorithms \citep{Friedman2001}, in  component-wise gradient boosting each base-learner only accounts for a single covariate. 
The model is hence exclusively updated with respect to the covariate that adds the most information. 
As a result, early stopping induces an intrinsic variable selection. \\ 
Together with the step length $\nu$, the stopping iteration $m_\text{stop}$ is thus the tuning parameter that determines the sparsity and degree of regularization of the model. 
Following \cite{Buehlmann2007}, who reasoned that the step length is not important as long as chosen small enough, $\nu$ is often set to a small constant, e.g. 0.1, which leaves $m_\text{stop}$ as the main tuning parameter. 
In order to determine $m_\text{stop}$, e.g. information criteria or resampling methods like cross-validation can be applied \citep{Buehlmann2007}. \\

Boosting GAMLSS differs from boosting models with a single predictor in that $K$ different submodels have to be fitted instead of one, where in addition to a larger overall model and variable selection also the relative complexity of the different submodels has to be determined. 
There are different options for that within the boosting framework. 
Originally, \cite{Mayr2012} proposed to update the predictors cyclically, in which case every predictor is updated in every iteration. 
In cyclical boosting algorithms, an individual stopping iteration has to be determined for every submodel, where the different stopping iterations are not independent and have to be optimized jointly, e.g.\ by grid search. \\ 
{In order to reduce the computational complexity of the tuning procedure, \cite{Thomas2018} introduced a non-cyclical boosting algorithm, which we will focus on in this work. 
As opposed to the cyclical 
\parfillskip=0pt\par}

\vspace*{1em}
\begin{minipage}{3.3em}
$\phantom{.}$
\end{minipage}
\begin{minipage}{32.5em}
\begin{algorithm}[H]
\SetAlgoLined
\caption{Non-cyclical component-wise gradient boosting algorithm \newline \hspace*{5.6em} 
for GAMLSS}
Initialize the predictors $\boldsymbol{\eta}_{\theta_1}^{[0]}, ... \, , \boldsymbol{\eta}_{\theta_K}^{[0]}$ 
with offset values. \\
\For{$m=1$ \text{\rm{to}} $m_\text{\rm{stop}}$}{
\For{$k=1$ \text{\rm{to}} $K$}{
\begin{enumerate}
\setlength\itemindent{-1em}
\item Compute the estimated negative gradient vector $\boldsymbol{u}^{[m]}_k$.
\item Fit every base-learner to the negative gradient vector:
\hspace*{3em} $\boldsymbol{u}_k^{[m]} \rightarrow \boldsymbol{h}^{[m]}_{j, \theta_k} \text{ for } j = 1, ... \, , J_k$ 
\item Select the best-fitting base-learner $\boldsymbol{h}^{[m]}_{j_k^*, \theta_k}$, where 
\hspace*{1.9em}  $j_k^* = \underset{j \in \{1, ..., J_k\}}{\text{arg min}} \sum_{i=1}^n \left(u_{k,i}^{[m]} - h^{[m]}_{j, \theta_k, i}\right)^2$. 
\item Choose the step length $\nu_{\theta_k}^{[m]}$.
E.g., $\nu_{\theta_k}^{[m]} = 0.1$. 
\item Compute the potential update $\boldsymbol{\eta}_{\theta_k}^{[m-1]} + \nu_{\theta_k}^{[m]} \cdot \boldsymbol{h}^{[m]}_{j_k^*, \theta_k}$ 
and the \\ \hspace*{-1.22em}  value of the loss function \\[0.5em] 
\hspace*{1.5em} $\rho_k^{[m]} = \rho \left(\boldsymbol{y}, \boldsymbol{\eta}_{\theta_k}^{[m-1]} + \nu_{\theta_k}^{[m]} \cdot \boldsymbol{h}^{[m]}_{j_k^*, \theta_k}, \boldsymbol{\eta}_{-\theta_k}^{[m-1]} \right)$.
\end{enumerate}
}
\vspace*{-0.3em}
\begin{enumerate}
\setcounter{enumi}{5}
\setlength\itemindent{-1em}
\item Determine the predictor $\boldsymbol{\eta}^{[m]}_{\theta_{k^*}}$, where $k^* = \underset{k \in \{1, ... , K\}}{\text{arg min}} \rho_k^{[m]}$. 
\item Update the predictors
$\boldsymbol{\eta}^{[m]}_{\theta_{k}} = 
\begin{cases}
\boldsymbol{\eta}_{\theta_k}^{[m-1]} +
\nu_{\theta_k}^{[m]} \cdot \boldsymbol{h}^{[m]}_{j_k^*, \theta_k} \hspace{0.5em} \text{ if } k = k^* \\
\boldsymbol{\eta}_{\theta_k}^{[m-1]} \hspace{6.8em} \text{ if } k \neq k^*
\end{cases}$ \hspace*{-1em}.
\end{enumerate}
}
\label{daub:algo_non-cyclical_boosting_GAMLSS}
\end{algorithm} 
\end{minipage}
 
\noindent updating scheme, only the predictor $\boldsymbol{\eta}_{\theta_{k^*}}$ whose update yields the largest improvement with respect to the loss function $\rho(\cdot)$ is updated in non-cyclical boosting algorithms. 
For $\boldsymbol{\eta}_{\theta_{k^*}}$ therefore holds
\begin{align*}
k^* = \underset{k \in \{1, ..., K\}}{\text{arg min}} \,
\rho\left(
\boldsymbol{y}, 
\boldsymbol{\eta}_{\theta_k}^{[m-1]} + \nu_{\theta_k}^{[m]} \cdot \boldsymbol{h}^{[m]}_{j_k^*, \theta_k}, 
\boldsymbol{\eta}_{-\theta_k}^{[m-1]}
\right) \, , 
\end{align*} 
where $\boldsymbol{h}^{[m]}_{j_k^*, \theta_k}$ is the best-performing fitted base-learner for $\theta_k$ in iteration $m$ and $\boldsymbol{\eta}_{-\theta_k}^{[m-1]}$ represents the set of predictors without $\boldsymbol{\eta}_{\theta_k}^{[m-1]}$, i.e. $\boldsymbol{\eta}_{-\theta_k}^{[m-1]} = \left\{\boldsymbol{\eta}_{\theta_l}^{[m-1]} \big\vert \, l \in \{1, ..., K\} \setminus \{k\} \right\}$. The whole non-cyclical boosting algorithm for GAMLSS is displayed in Algorithm~\ref{daub:algo_non-cyclical_boosting_GAMLSS}, where we on purpose leave the step length in step 4 more flexible than originally proposed by \cite{Thomas2018}. 
\cite{Thomas2018} set $\nu_{\theta_k}^{[m]}$ to the same small fixed value $\bar{\nu}$ for all distribution parameters and iterations, e.g. $\bar{\nu} = 0.1$. \\ 
While the cyclical boosting algorithm outsources the task of finding the relative complexity of the different submodels to the tuning procedure, with a non-cyclical updating scheme it is determined intrinsically based on the comparison of the potential updates' reduction of the loss function. 
That has two main implications.
On the one hand, the tuning procedure is reduced to a one-dimensional optimization problem avoiding the need of a computationally expensive grid search.
On the other hand, Thomas et al.'s (2018) non-cyclical boosting algorithm can result in an imbalanced overall model as \cite{Zhang2022} showed exemplarily for a Gaussian location and scale model with a large variance. 
This issue will be discussed in more detail in section~\ref{daub:subsection_BLratio_steplengths}. 
Moreover, in these cases many iterations are needed until stopping, which results in a long run time. 
Both limitations were addressed by \cite{Zhang2022} by using an adaptive step length, which will be elaborated on in the following.

\subsection{Adaptive step lengths in non-cyclical boosting algorithms for GAMLSS}
\label{daub:section_nonclyclical_boosting_with_adaptive_steplengths}

\cite{Zhang2022} proposed to use different step lengths for different predictors as well as 
over the iterations. 
For an update of $\boldsymbol{\eta}_{\theta_k}$ in iteration $m$, they specified the adaptive step length $\nu^{[m]}_{\theta_k}$ in step 4 of the non-cyclical boosting algorithm (see Algorithm~\ref{daub:algo_non-cyclical_boosting_GAMLSS}) as 
\begin{align}
\nu^{[m]}_{\theta_k} = \lambda_s \cdot \nu^{*[m]}_{\theta_k} \, \, \text{with } \,
\nu_{\theta_k}^{*[m]} = \underset{\nu}{\text{arg min}} \, 
\rho \,	\left(\boldsymbol{y}, \boldsymbol{\eta}_{\theta_k}^{[m-1]} + \nu \cdot \boldsymbol{h}^{[m]}_{j_k^*, \theta_k}, \boldsymbol{\eta}_{-\theta_k}^{[m-1]} \right) \, ,
\label{daub:formula_shrunk_optimal_steplength}
\end{align} 
where $\lambda_s$ is a shrinkage factor, for example 0.1. 
The step length defined in (\ref{daub:formula_shrunk_optimal_steplength}) thus reflects how much the update in question could potentially reduce the loss function,
while the shrinkage factor ensures that the updates are small in order to prevent overfitting \citep{Hepp2016}. 
In this step length definition, \cite{Zhang2022} took up Friedman's (2001) idea of using shrunk optimal step lengths from the initial statistical boosting algorithm and demonstrated exemplarily for a Gaussian response variable that this adaptive step length ensures a natural balance in the updates of the different submodels.
Since it moreover prevents late stopping, it addresses the limitations of Thomas et al.'s (2018) non-cyclical boosting algorithm outlined in the previous section. \\
\cite{Zhang2022} introduced two options for determining the proposed adaptive step lengths. 
Firstly, the optimization problem in (\ref{daub:formula_shrunk_optimal_steplength}) can be solved numerically via line search.
Secondly, an analytical solution for $\nu_\mu^*$ as well as a convergence result for $\nu_\sigma^*$ was derived for the special case of a Gaussian response variable.
In order to generalize this concept to non-Gaussian GAMLSS, an optimal step length would either have to be derived for every submodel of every response variable separately, where in many cases an analytical closed form solution does not exist, or a line search could be conducted.
A line search is however computationally more demanding and requires a prespecified search interval. 
We therefore propose a third option for obtaining adaptive step lengths, which is based on the algorithm's intrinsic information, more precisely on the size of the different base-learners.

\subsubsection{Adaptive step lengths based on the ratio of base-learner norms}
\label{daub:subsection_BLratio_steplengths}

\begin{figure}[t]
\centering 
\includegraphics[width=0.67\textwidth]{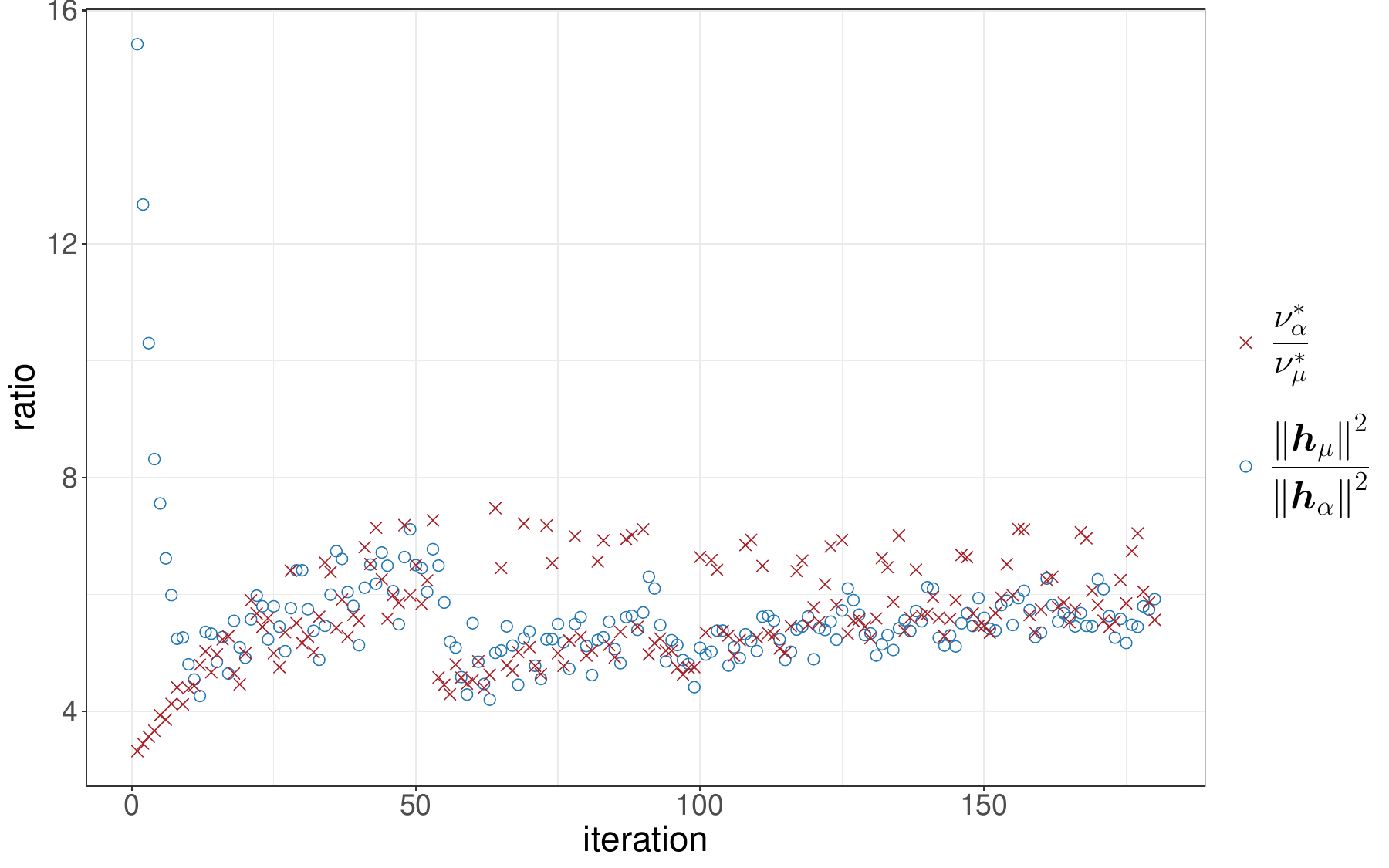}
\caption{\label{daub:fig_ratios_steplengths_BL} Inverse ratio of numerically obtained optimal step lengths and ratio of squared Euclidean base-learner norms for an exemplary simulation run in the negative binomial simulation setting (\ref{daub:formula_simulation_model_negbinom}) without additional non-informative covariates. }
\end{figure}

As alternative adaptive step length scheme, we propose to define the step length $\nu_{\theta_l}^{[m]}$ for updating $\boldsymbol{\eta}_{\theta_l}^{[m-1]}$, $l \in \{1, ..., K\} \setminus \{k\}$, in iteration $m$ as the step length of a reference parameter $\theta_k$ rescaled by the ratio of norms of the respective fitted base-learners, i.e.
\begin{align}
\nu_{\theta_l}^{[m]} = \nu_{\theta_k}^{[m]} \,
\frac{\left\Vert \boldsymbol{h}_{j_k^*, \theta_k}^{[m]} \right\Vert_2^2}
{\left\Vert \boldsymbol{h}_{j_l^*, \theta_l}^{[m]} \right\Vert_2^2} \, ,
\label{daub:formula_BL_steplengths}
\end{align}
where $\lVert \boldsymbol{x} \rVert_2$ denotes the Euclidean norm of $\boldsymbol{x} \in \mathbb{R}^n$. \\
This construction and especially the choice of the squared Euclidean norm as the measure for the length of a vector is motivated by the observation that the ratio of optimal step lengths $\nu_{\theta_l}^{*[m]} / \nu_{\theta_k}^{*[m]}$ behaves approximately inverse to the ratio of squared Euclidean base-learner norms in the considered cases, see for example Figure~\ref{daub:fig_ratios_steplengths_BL}.
Even though in most cases the base-learner ratio and the inverse ratio of optimal step lengths do not coincide right from the start, they converge quickly and then stay on a similar level, where however differences between step length ratios of different covariates in the same predictor cannot always be captured.
For the shrunk optimal step lengths $\nu^{[m]}_{\theta_k}$ and $\nu^{[m]}_{\theta_l}$ 
then holds
\begin{align}
\frac{\nu_{\theta_l}^{[m]}}{\nu_{\theta_k}^{[m]}} 
= \frac{\nu_{\theta_l}^{*[m]}}{\nu_{\theta_k}^{*[m]}} 
\approx 
\frac{\left\Vert \boldsymbol{h}_{j_k^*, \theta_k}^{[m]} \right\Vert_2^2}
{\left\Vert \boldsymbol{h}_{j_l^*, \theta_l}^{[m]} \right\Vert_2^2}
\label{daub:formula_ratios_invSL_BL}
\end{align}
due to the multiplicative structure of shrunk optimal step lengths (see (\ref{daub:formula_shrunk_optimal_steplength})).\\ 
Defining the size of the potential update of $\boldsymbol{\eta}^{[m-1]}_{\theta_k}$ 
by $ \zeta_{\theta_k}^{[m]} := \nu_{\theta_k}^{[m]} \, \left\Vert \boldsymbol{h}_{j_k^*, \theta_k}^{[m]} \right\Vert_2^2$, the approximately inverse relationship of base-learner norms and step lengths in (\ref{daub:formula_ratios_invSL_BL}) can be rewritten as
\begin{align}
\zeta_{\theta_k}^{[m]} 
= \nu_{\theta_k}^{[m]} \,\left\Vert \boldsymbol{h}_{j_k^*, \theta_k}^{[m]} \right\Vert_2^2 
\approx \nu_{\theta_l}^{[m]} \, \left\Vert \boldsymbol{h}_{j_l^*, \theta_l}^{[m]} \right\Vert_2^2 
= \zeta_{\theta_l}^{[m]} \, . 
\label{daub:formula_stepsizes_approx}
\end{align}
Please be aware of the terminological difference between the step length $\nu_{\theta_k}^{[m]}$,
which is the scaling factor of the fitted base-learner, and the update size $\zeta_{\theta_k}^{[m]}$, which corresponds to the product of step length and base-learner norm.
The update sizes of an exemplary simulation run with shrunk optimal step lengths are displayed in Appendix~\ref{daub:appendix_simu_negbinom}, Figure~\ref{daub:fig_stepsizes_single_run_negbinom}. \\

Equation (\ref{daub:formula_stepsizes_approx}) illustrates how shrunk optimal step lengths achieve a natural balance in the overall model. 
In order to gain further insight into the solution, we first take a closer look at the problem. \\
When applying a non-cyclical boosting algorithm to data for which the negative gradients live on different scales, the scale difference carries over to the fitted base-learners. 
Since with fixed step lengths the potential predictor updates equal the fitted base-learners scaled by the same constant, the non-cyclical boosting algorithm then compares potential updates of very different sizes with respect to their improvement of the loss function in order to decide which predictor is updated (see Algorithm~\ref{daub:algo_non-cyclical_boosting_GAMLSS}, step 6). 
The negative gradients' size difference therefore results in a severe structural advantage for the predictor with the larger negative gradient. 
For example for a Gaussian location and scale model with a large variance, the negative gradients and thus the fitted base-learners for $\eta_\mu$ are much smaller than the negative gradients and fitted base-learners for $\eta_\sigma$. 
Therefore, $\eta_\sigma$ is updated in almost every iteration in the beginning until its paths have almost converged, which leads to $\eta_\sigma$ being overfitted or, depending on the stopping iteration, $\eta_\mu$ being underfitted. \\
Instead of carrying the negative gradients' scales along until the predictor update, optimal step lengths compensate for structurally large or small base-learners such that the potential predictor updates are approximately equally sized (see (\ref{daub:formula_stepsizes_approx})). 
Therefore, a fairer decision of which predictor to update can be achieved resulting in a more balanced overall model in the end. 
Base-learner based step lengths adopt this property in order to yield a fair update selection and by definition fix the sizes of all potential updates to the size of the reference parameter's update.

\subsubsection{Reference step lengths}

When applying base-learner based step lengths, a reference step length, i.e. the step length scheme of the reference parameter, has to always be specified additionally. 
One option is to use a shrunk optimal step length, which has the advantages that all updates automatically have a sensible size and overly long run times can be avoided.   
With a shrunk optimal step length as reference step length, the base-learner based step lengths can be understood as an approximation of the shrunk optimal step lengths corresponding to the other predictors. 
In order to determine the shrunk optimal step length of the reference parameter, either an analytically derived solution or a numerical method can be used. 
When considering a negative binomial or Weibull distributed response variable, a closed form analytical term that approximates the optimal step length and can serve as reference step length can be derived, which we will elaborate on in section~\ref{daub:subsection_anal_approx_opt_steplength}. \\

Alternatively, a fixed step length can serve as reference step length.
In that case, all potential updates in iteration $m$ are set to the size $\zeta_{\theta_k}^{[m]} $ of the reference parameter's potential update that is obtained with the same step length in all iterations $m \in \{1, ..., m_\text{stop}\}$. 
By combining fixed and base-learner based step lengths, the characteristic of optimal step lengths that addresses the imbalancedness problem is transferred to the fixed step length approach. 
As will be shown in section~\ref{daub:section_simulations}, these step lengths yield very similar estimation results to base-learner based step lengths with a shrunk optimal reference step length. \\
Using a fixed step length as reference step length however has the following drawback.
Unlike shrunk optimal step lengths, 
base-learner based step lengths with a fixed reference step length do not ensure updates of reasonable size and can thus potentially result in very long run times.
With a structurally very small negative gradient as reference, base-learner based step lengths actually amplify the problem since all updates are set to that very small update size.
When using a fixed reference step length, we therefore recommend to choose the predictor as reference that is updated more frequently in the beginning of a fixed step length approach in order to avoid a problem with very small updates. 
Alternatively, this issue could be addressed by choosing a large fixed reference step length. 
While that is common practice for boosting algorithms with one predictor, finding a reasonable fixed reference step length can be more difficult when boosting GAMLSS as in some cases the optimal step lengths are very large \citep{Zhang2022}.

\subsubsection{Optimal step lengths in negative binomial and Weibull models}
\label{daub:subsection_anal_approx_opt_steplength}

In order to implement base-learner based step lengths with a shrunk optimal reference step length efficiently without numerical optimization, we next present an analytical approximate solution of (\ref{daub:formula_shrunk_optimal_steplength}) for $\mu$ in a negative binomial location and scale model as well as for $\lambda$ in a Weibull scale and shape model.
The models are chosen as prominent examples of GAMLSS for count and survival data, respectively, 
and the distributional parameter that is typically of higher interest is used as the reference parameter. \\

First, we consider
\begin{align*}
y_i \sim \mathcal{NB}(\mu_i, \alpha_i) ,
\end{align*} 
where $\mu_i$ is the mean, $\alpha_i$ the overdispersion parameter and the variance of $y_i$ is $\mu_i + \alpha_i \mu_i^2$. 
The corresponding probability mass function is 
\begin{align}
f(y_i;\mu_i, \alpha_i) 
= \frac{\Gamma(y_i + 1/\alpha_i)}{\Gamma(y_i+1) \, \Gamma(1/\alpha_i)}
\left( \frac{1}{1+\alpha_i \mu_i}\right)^{\frac{1}{\alpha_i}}
\left( 1 - \frac{1}{1+\alpha_i \mu_i} \right)^{y_i} \, .
\label{daub:formula_pmf_neg_binom}
\end{align}
This formulation of the negative binomial distribution follows \cite{Hilbe2011}, who refers to it as NB2 model, and is implemented as \texttt{NBI} in the gamlss package in R \citep{Stasinopoulos2022}. 
Following common practice, logarithmic link functions are used for both distribution parameters \citep{Hilbe2011}; we therefore have
\begin{align*}
\mu_i = \exp(\eta_{\mu, i}) \quad \text{and} \quad \alpha_i = \exp(\eta_{\alpha, i}) \, .
\end{align*}
In order to determine the optimal step length $\nu_\mu^{*[m]}$ in iteration $m$, the following optimization problem with the negative log-likelihood $-\ell(\cdot)$ as the loss function has to be solved
\begin{align*}
\nu^{* [m]}_\mu = \underset{\nu_\mu}{\text{arg min}} \, -\ell &\left(\exp\left(\boldsymbol{\eta}_\mu^{[m-1]} + \nu_\mu \, \boldsymbol{h}^{[m]}_{j_\mu^*, \mu} \right), \boldsymbol{\alpha}^{[m-1]}; \boldsymbol{y} \right) \, ,
\end{align*}
where $\exp(\cdot)$ is applied element-wise.
Denoting the $i$th component of $\boldsymbol{h}^{[m]}_{j_\mu^*, \mu}$ by $h_{\mu,i}^{*[m]}$ for notational convenience, the first order condition for the optimal step length is 
\begin{align}
0 = - \frac{\partial \ell}{\partial \nu_\mu} \Bigg\vert_{\nu_\mu=\nu_\mu^{*[m]}}
= - \sum_{i=1}^n \frac{h_{\mu,i}^{*[m]} \left(y_i - \exp\left(\eta_{\mu,i}^{[m-1]} + \nu_\mu^{*[m]} h_{\mu,i}^{*[m]}\right)\right)}{1 + \alpha_i^{[m-1]} \exp\left(\eta_{\mu,i}^{[m-1]} + \nu_\mu^{*[m]} h_{\mu,i}^{*[m]}\right)} \, .
\label{daub:formula_1st_order_condition_negbinom}
\end{align}
Equation (\ref{daub:formula_1st_order_condition_negbinom}) cannot be solved for $\nu_\mu^{*[m]}$ in closed form. 
In order to nevertheless obtain a closed form representation of the optimal step length, we use the following two approximations. 
First, we approximate $\nu_\mu^{*[m]}$ in the denominator by the optimal step length used in the last (potential) update of $\boldsymbol{\eta}_\mu$ with respect to the $j_\mu^*$th covariate, $\nu_\mu^{[\text{prev}, {j^*}]}$. 
For the first (potential) update of $\boldsymbol{\eta}_\mu$ with respect to the $j_\mu^*$th covariate, the optimal step length from the previous iteration is considered instead and in the first iteration $\nu_\mu^{*[m]}$ is obtained via line search.
The reason for approximating $\nu_\mu^{*[m]}$ by the optimal step length of a previous iteration is that the step lengths typically do not fluctuate much and we thus expect the new step length to be similar to the previous step length. \\
Based on the same reasoning, we approximate the term $\exp\left(\eta_{\mu,i}^{[m-1]} + \nu_\mu^{*[m]} h_{\mu,i}^{*[m]}\right)$ in the numerator via a Taylor polynomial of degree 1 around $\nu_\mu^{[\text{prev}, {j^*}]}$, i.e.
\begin{align*}
\exp\left(\eta_{\mu,i}^{[m-1]} + \nu_\mu^{*[m]} h_{\mu,i}^{*[m]}\right)  
\approx \exp\left(\eta_{\mu,i}^{[m-1]}\, + \, \nu_\mu^{[\text{prev}, j]} \, h_{\mu,i}^{*[m]}\right) \, \cdot
\left[ 1 + h_{\mu,i}^{*[m]} \, \left(\nu_\mu^{*[m]} - \nu_\mu^{[\text{prev}, j]}\right) \right] \, .
\end{align*}
Applying both approximations, the first order condition from (\ref{daub:formula_1st_order_condition_negbinom}) becomes
\begin{align*}
0 \approx - \sum_{i=1}^n \frac{h_{\mu,i}^{*[m]} \left(y_i - \left(\exp\left(\eta_{\mu,i}^{[m-1]}\, + \, \nu_\mu^{[\text{prev}, j]} \, h_{\mu,i}^{*[m]}\right) \, \cdot
\left[ 1 + h_{\mu,i}^{*[m]} \, \left(\nu_\mu^{*[m]} - \nu_\mu^{[\text{prev}, j]}\right) \right]\right)\right)}{1 + \alpha_i^{[m-1]} \exp\left(\eta_{\mu,i}^{[m-1]} + \nu_\mu^{[\text{prev}, j^*]} h_{\mu,i}^{*[m]}\right)}  \, .
\end{align*}
\begin{figure}[b!]
\center
\includegraphics[width=0.95\textwidth]{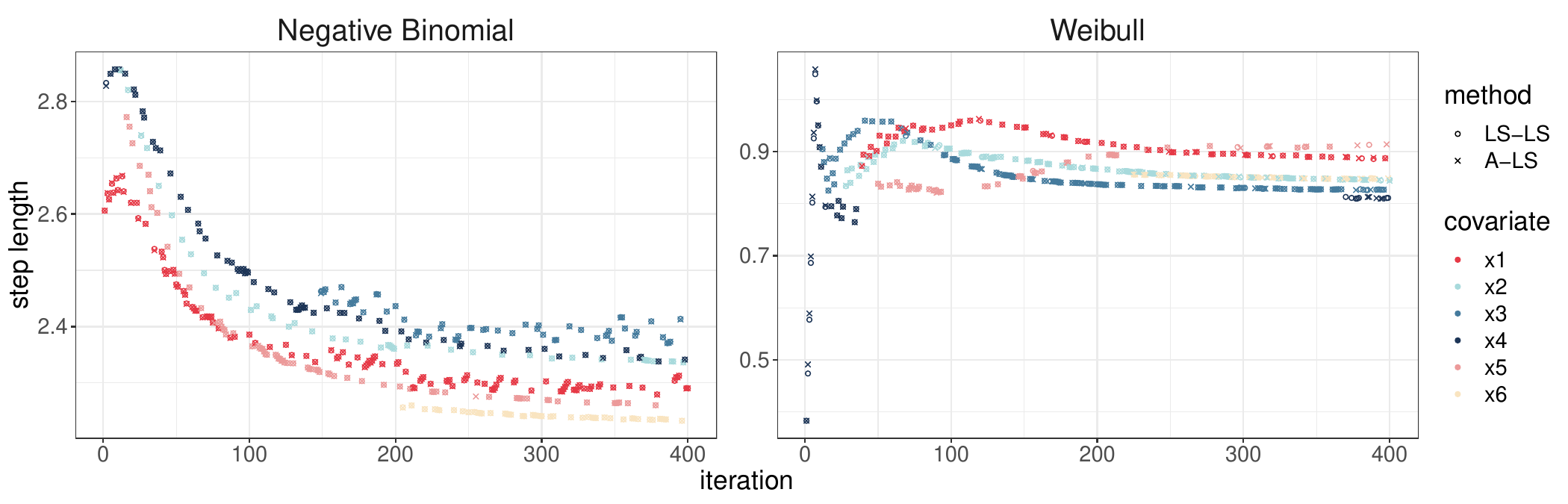}
\caption{\label{daub:fig_steplengths_analytical} Optimal step lengths $\nu_\mu^{*[m]}$ and $\nu_\lambda^{*[m]}$ obtained numerically (LS-LS) and via the analytical approximations (A-LS) for an exemplary simulation run without additional non-informative covariates in the negative binomial simulation setting (\ref{daub:formula_simulation_model_negbinom}) and the Weibull simulation setting (\ref{daub:formula_simulation_model_weibull}), respectively.
Further explanations on the abbreviations of the step length approaches can be found in Figure~\ref{daub:fig_overview_steplengths} or in section~\ref{daub:subsection_simulation_gaussian}.}
\end{figure} 
We therefore obtain an approximated optimal step length $\nu_\mu^{*[m]}$ in a negative binomial location and scale model of
\begin{align}
\nu_\mu^{*[m]} \approx 
\frac{\sum_{i=1}^n \frac{h_{\mu,i}^{*[m]} \left(y_i - \left(\exp\left(\eta_{\mu,i}^{[m-1]}\, + \, \nu_\mu^{[\text{prev}, j]} \, h_{\mu,i}^{*[m]}\right) \, \cdot
\left[ 1 - h_{\mu,i}^{*[m]} \, \nu_\mu^{[\text{prev}, j]} \right]\right)\right)}{1 + \alpha_i^{[m-1]} \exp\left(\eta_{\mu,i}^{[m-1]} + \nu_\mu^{[\text{prev}, j^*]} h_{\mu,i}^{*[m]}\right)}}
{\sum_{i=1}^n \frac{{h_{\mu,i}^{*[m]}}^2 \exp\left(\eta_{\mu,i}^{[m-1]}\, + \, \nu_\mu^{[\text{prev}, j]} \, h_{\mu,i}^{*[m]}\right)}{1 + \alpha_i^{[m-1]} \exp\left(\eta_{\mu,i}^{[m-1]} + \nu_\mu^{[\text{prev}, j^*]} h_{\mu,i}^{*[m]}\right)}} \, .
\label{daub:formula_opt_step_length_mu_neg_binom}
\end{align}
For a more detailed derivation, see Appendix~\ref{daub:appendix_opt_steplength_neg_binom}. \\

$\nu_\lambda^{*[m]}$ in a Weibull scale and shape model can be approximated using the same concepts, which yields 
\begin{small}
\begin{align*}
\nu_\lambda^{*[m]} \approx 
- \frac{\sum_{i=1}^n  \, h_{\lambda,i}^{*[m]} \, k^{[m-1]}_i
- \sum_{i=1}^n  \,  h_{\lambda,i}^{*[m]} \, k^{[m-1]}_i \, y_i^{k^{[m-1]}_i} 
\exp\left(\eta_{\lambda,i}^{[m-1]} \, + \, \nu_\lambda^{[\text{prev}, j]} \, h_{\lambda,i}^{*[m]}\right)^{-k^{[m-1]}_i} \hspace*{-0.2em}
\left[ 1 + k^{[m-1]}_i \, h_{\lambda,i}^{*[m]} \, \nu_\lambda^{[\text{prev}, j]} \right]}
{\sum_{i=1}^n  \, {h_{\lambda,i}^{*[m]}}^2 \, {k^{[m-1]}_i}^2 \, y_i^{k^{[m-1]}_i} \, \exp\left(\eta_{\lambda,i}^{[m-1]} \, + \, \nu_\lambda^{[\text{prev}, j]} \, h_{\lambda,i}^{*[m]}\right)^{-k^{[m-1]}_i}}
\end{align*}
\end{small}
A graphical comparison between the analytically approximated and numerically obtained optimal step lengths in an exemplary simulation run is depicted in Figure~\ref{daub:fig_steplengths_analytical} and a detailed derivation of the approximated $\nu_\lambda^{*[m]}$ can be found in Appendix \ref{daub:appendix_opt_steplength_weibull}. \\

\begin{figure}[htb!]
\center
\includegraphics[width=0.95\textwidth]{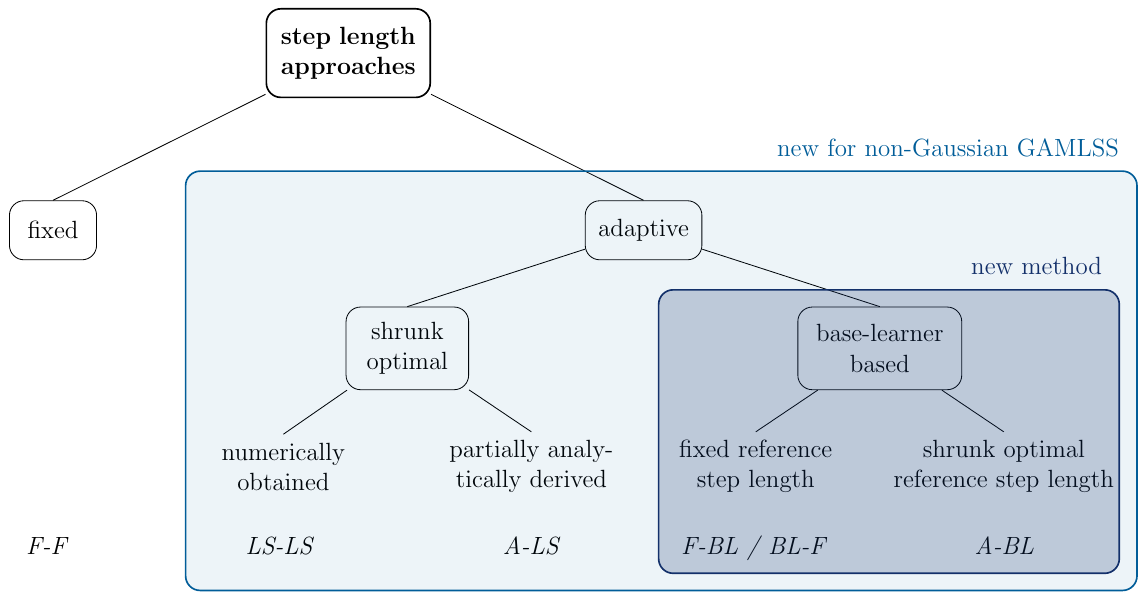}
\caption{\label{daub:fig_overview_steplengths} Overview of the different step length approaches.}
\end{figure}

An overview of the different step length approaches is displayed below (see Figure~\ref{daub:fig_overview_steplengths}). \\ 

%% file: simus.tex
\section{Simulation Study}
\label{daub:section_simulations}

In the following, we investigate the performance of the newly proposed adaptive step length scheme by means of a simulation study. 
We first demonstrate for a Gaussian response variable that our base-learner based step lengths yield competitive results to a non-cyclical boosting algorithm with shrunk optimal step lengths. 
Note that we are not trying to outperform Zhang et al.'s (2022) shrunk optimal approach but the goal is to show that using the new adaptive step lengths yields similar results while being computationally more efficient.
In a second simulation setting, we consider negative binomial and Weibull distributed response variables and examine the performance of the different adaptive step length approaches compared to Thomas et al.'s (2018) non-cyclical boosting algorithm with fixed step lengths. 
In addition to an investigation of the performance of the base-learner based approach, we want to demonstrate that both adaptive step length approaches yield a more balanced overall model than fixed step lengths in this setting as well as that the proposed analytical approximations of the optimal step lengths are sufficiently accurate.

\subsection{Gaussian location and scale model}
\label{daub:subsection_simulation_gaussian}

In order to investigate if base-learner based step lengths have a similar performance as shrunk optimal step lengths, we consider the simulation setup of \cite{Zhang2022} with a lower intercept in $\eta_\sigma$ and hence a moderately high variance. 
Specifically, $n=500$ observations of the response variable $y_i$ are sampled from $\mathcal{N}(\mu_i, \sigma_i)$ with
\begin{align}
\eta_{\mu, i} &= \mu_i = x_{1i} + 2 \cdot x_{2i} + 0.5 \cdot x_{3i} - x_{4i} \nonumber \\
\eta_{\sigma, i} &= \log(\sigma_i) = 2 + 0.2 \cdot x_{3i} + 0.1 \cdot x_{4i} - 0.1 \cdot x_{5i} - 0.2 \cdot x_{6i} ,
\label{daub:formula_simulation_model_Gaussian}
\end{align}
where $x_1,...,x_6$ are drawn independently from a uniform distribution on $[-1,1]$. 
In this simulation setting, thus only the covariates $x_{3}$ and $x_{4}$ are shared between both predictors
and the three cases of 0, 10 and 150 additional non-informative covariates are considered. \\ 
We apply the non-cyclical boosting algorithm with four different step length schemes. Two step length approaches with analytically derived $\nu_\mu$, where the step length for an update of $\eta_\sigma$ is either obtained by line search (A-LS) or based on the base-learner based approach (A-BL) are considered.
Note that the abbreviations of the step length schemes are combined of the step length approach applied to obtain $\nu_\mu$ (first component) and the approach for $\nu_\sigma$ (second component).
Additionally, a fixed step length of 0.1 for $\sigma$ is used combined with a base-learner based step length for $\mu$ (BL-F) as well as a fixed step length of 0.1 for $\mu$ (F-F).
For computing shrunk optimal step lengths, a shrinkage factor of $\lambda_s=0.1$ is considered and, if necessary, a search interval of $[0,10]$.
The algorithm is stopped early at an iteration obtained via 10-fold cross-validation and $B = 100$ simulation runs are conducted. \\

\begin{figure}[b!]
\centering
\includegraphics[width= 0.9\textwidth]{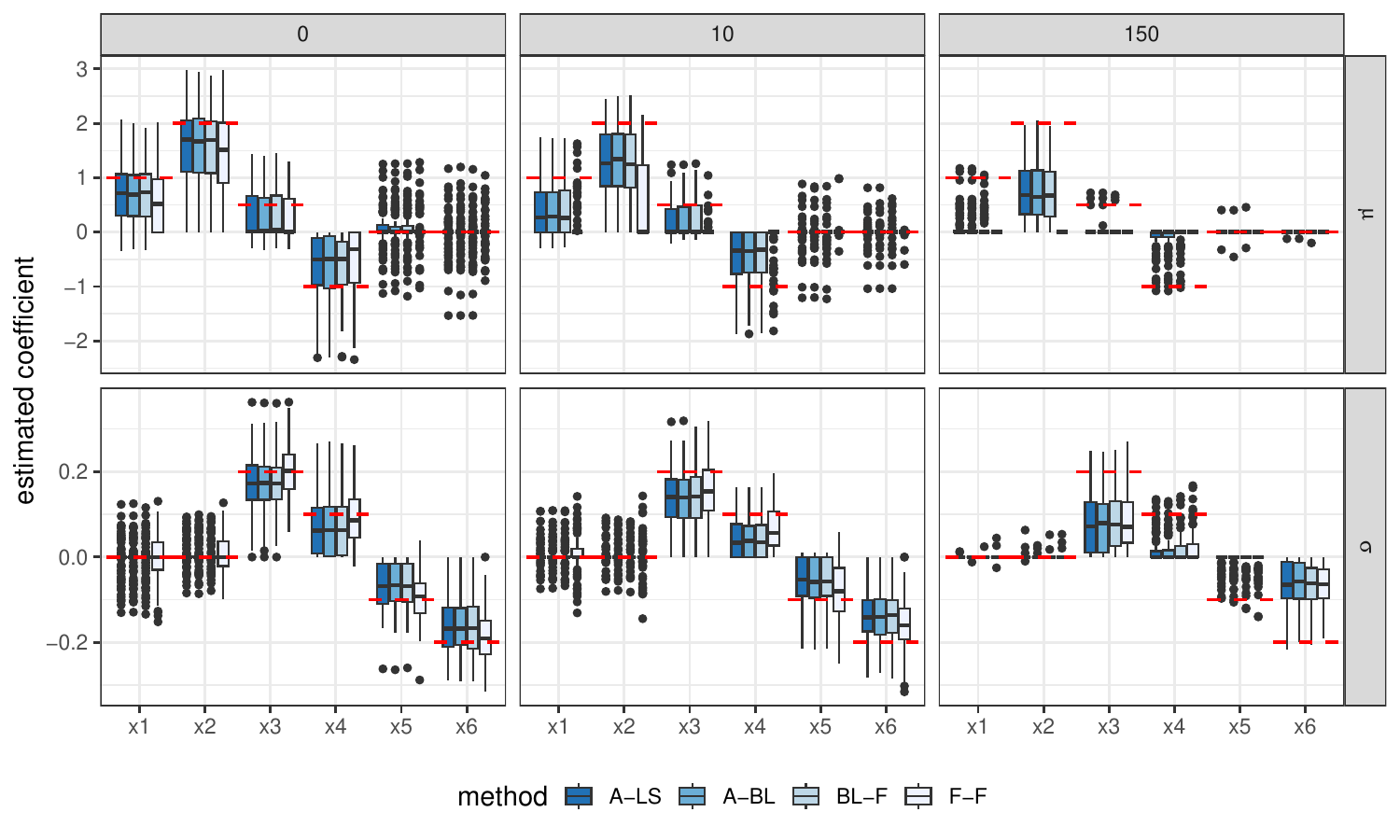}
\caption{\label{daub:fig_overview_coeffs_Gaussian} Distribution of the coefficient estimates in the Gaussian simulation setting (\ref{daub:formula_simulation_model_Gaussian}) for a varying number of additional non-informative covariates (columns).
The red dashed lines represent the true coefficients.} 
\end{figure}

In order to investigate the performance of the new adaptive step lengths, we compare the coefficient estimates of the base-learner based approaches A-BL and BL-F with the ones obtained using shrunk optimal step lengths (A-LS) on the one hand and fixed step lengths (F-F) on the other.
In addition, variable selection and balancedness measures are included in the comparison. \\
From the coefficients' distributions it is evident that a fixed step length approach yields a more pronounced coefficient shrinkage of $\eta_\mu$ than of $\eta_\sigma$, which is enhanced in settings with 10 or more additional non-informative covariates (see Figure~\ref{daub:fig_overview_coeffs_Gaussian}). 
In contrast to that, the coefficients of the different predictors are shrunk by a similar degree in the adaptive step length approaches thus resulting in a more balanced overall model. 
Due to the high variance in this simulation setup, the variation of the estimated coefficients is quite high in all approaches. \\
These findings on the model's balancedness are further supported when considering the number of covariates selected for $\eta_\mu$ relative to the number of covariates in $\eta_\sigma$.
Since in the true model we have four informative covariates in each predictor (see (\ref{daub:formula_simulation_model_Gaussian})), 
the ratio of the number of covariates included in $\eta_\mu$ relative to $\eta_\sigma$ (SCR$_{\mu/\sigma}$) should be around 1 in a balanced overall model. 
When comparing the different step length approaches with respect to this balancedness measure in a setup without additional non-informative covariates, we find that the three adaptive step length approaches have an SCR$_{\mu/\sigma}$ of 1 in median, while the fixed step length approach selects about twice as many covariates for $\eta_\sigma$ compared to $\eta_\mu$ in median, which corresponds to a median SCR$_{\mu/\sigma}$ of 0.5. 
With an increasing number of additional non-informative covariates, the imbalancedness in the overall model resulting from the fixed step length scheme is enhanced. 
For example in a setup with 10 additional non-informative covariates, in over 50\% of the cases the model is not updated with respect to $\eta_\mu$ at all, whereas in median 9 covariates are included in $\eta_\sigma$.
The three adaptive step length approaches on the other hand have a median ratio of number of selected covariates of about 0.83.
For an overview of the distribution of the SCR$_{\mu/\sigma}$ for different numbers of additional non-informative covariates, see Figure~\ref{daub:fig_overview_selCov_ratio_Gaussian} in Appendix~\ref{daub:appendix_simu_gaussian}. \\
The imbalancedness of the fixed step length approach increasing in the number of additional non-informative covariates can be attributed to the fact that the non-cyclical boosting algorithm with fixed step lengths tends to almost exclusively update $\eta_\sigma$ in the first iterations here and then mainly update $\eta_\mu$ after the coefficients of $\eta_\sigma$ have almost converged (see section~\ref{daub:subsection_BLratio_steplengths} for a discussion on this issue). 
Since with more non-informative covariates the algorithm tends to stop earlier in order to prevent overfitting by non-informative covariates entering the model, the update order in the beginning has a stronger impact on the final model. 
These results are in line with Zhang et al.'s (2022) findings on Gaussian location and scale models with a large variance and show that the base-learner based step length approaches result in a similarly balanced overall model as shrunk optimal step lengths. \\

\begin{figure}[t!]
\includegraphics[width=\textwidth]{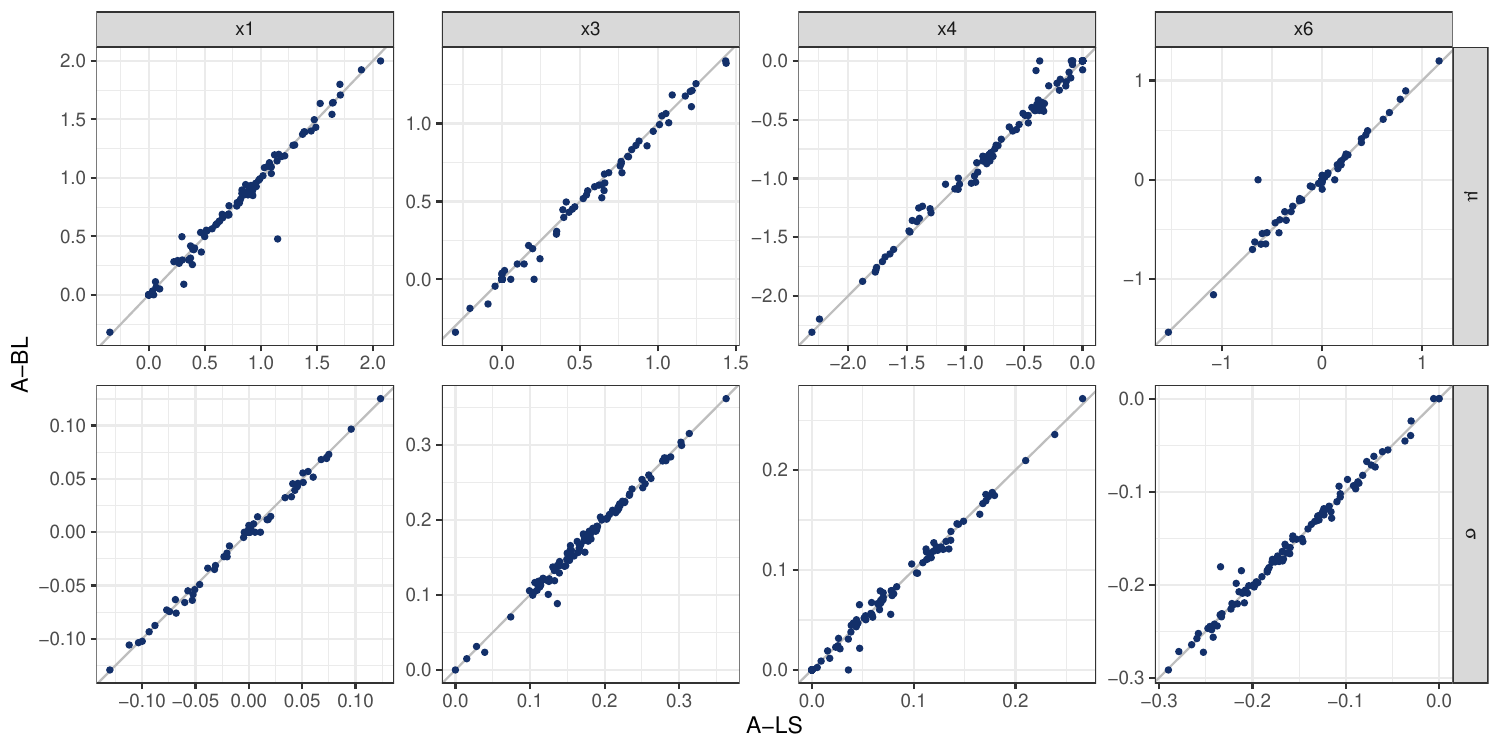}
\caption{\label{daub:fig_scatter_selCoefs_Gaussian} Direct comparison of the coefficient estimates of the two adaptive step length approaches A-LS and A-BL for the 100 simulation runs in the Gaussian simulation setting (\ref{daub:formula_simulation_model_Gaussian}) without additional non-informative covariates. }
\end{figure}

In order to compare the coefficient estimates of the adaptive step length approaches more closely, the coefficients of the different simulation runs are plotted against each other for selected covariates (see Figure~\ref{daub:fig_scatter_selCoefs_Gaussian} exemplarily for the coefficients corresponding to $x_1, x_3, x_4, x_6$ in the simulation setting without additional non-informative covariates).
The direct comparison of the coefficients supports the impression from above that A-BL and A-LS yield very similar coefficients. 
Comparing the coefficients of BL-F and A-LS shows a similar picture over all settings (see Figure~\ref{daub:fig_scatter_selCoefs_BLF_Gaussian} in Appendix~\ref{daub:appendix_simu_gaussian} exemplarily for the simulation setting  without additional non-informative covariates). 
In addition to the coefficient estimates, also the number of false positives and false negatives of the non-cyclical boosting algorithms with different adaptive step length approaches do not differ substantially, which indicates a similar variable selection behavior (see Figures~\ref{daub:fig_overview_FP_Gaussian} and \ref{daub:fig_overview_FN_Gaussian} in Appendix~\ref{daub:appendix_simu_gaussian}).
In accordance with the previous reasoning, the fixed step length approach has considerably more false negatives in $\eta_\mu$ and more false positives in $\eta_\sigma$ than the adaptive step length approaches and vice versa, again indicating that the fixed step length approach favors updates of $\eta_\sigma$ over updates of $\eta_\mu$ in this setup. \\

With the simulation results of the different adaptive step length approaches being very similar, it remains to be checked if the base-learner based approach improves the computational efficiency of the non-cyclical boosting algorithm.
When comparing the number of iterations needed until stopping, we find that A-LS and A-BL have a similar stopping iteration while BL-F stops after approximately half as many iterations in median and the fixed step length approach runs for considerably more iterations (see Table~\ref{daub:table_run_time_stopping_it_gaussian}). 
The lower stopping iteration of BL-F compared to A-BL can be attributed to the fact that a $\nu_\sigma$ of 0.1 is approximately twice the size of the shrunk optimal $\nu_\sigma$ in this case \citep[see][]{Zhang2022}. 
Conversely, the shrunk optimal step lengths for updates of $\mu$ are considerably higher than fixed step lengths of 0.1, which results in large stopping iterations for F-F.
With respect to the run time, 
using the base-learner based approach instead of numerically obtained shrunk optimal step lengths for $\nu_\sigma$ yields a reduction of about 10\% in median due to the omission of line search.
Note that the amount of run time reduction depends heavily on the chosen search interval 
as well as on the number of covariates that can potentially be included. 
For both BL-F and F-F, the stopping iteration drives the run time reduction and increase, respectively, compared to A-BL. \\

\begin{table}[t!] 
      \centering
      \begin{flushleft}
      \end{flushleft}
      \vspace*{-1em}
\begin{tabular}{lllll}
  \toprule[0.09 em]
 & \hspace*{0.6em} A-LS & \hspace*{0.6em} A-BL & \hspace*{0.6em} BL-F & \hspace*{1.2em} F-F \\ 
  \midrule
1st Qu. & 0.086 (68) & 0.078 (76) & 0.039 (38) & 0.924 (818) \\ 
  Median & 0.117 (102) & 0.107 (105) & 0.050 (50) & 2.059 (1,823) \\ 
  3rd Qu. & 0.181 (154) & 0.162 (168) & 0.078 (78) & 3.765 (3,122) \\ 
   \bottomrule[0.09 em]
\end{tabular}
\caption{\label{daub:table_run_time_stopping_it_gaussian}
Quartiles of the run times until $m_\text{stop}$ in seconds and stopping iterations (in parenthesis) in the Gaussian simulation setting (\ref{daub:formula_simulation_model_Gaussian}) without additional non-informative covariates.}
\end{table}

The simulation results for a Gaussian location and scale model thus indicate that the new base-learner based step lengths ensure a natural balance between the submodels. 
While moreover all adaptive step length approaches yield very similar estimation results, the base-learner based approaches outperform Zhang et al.'s (2022) shrunk optimal step lengths with respect to the run time.

\subsection{Non-Gaussian GAMLSS}
\label{daub:subsection_simulation_non-gaussian}

Next, we want to demonstrate that the proposed adaptive step length scheme can be applied for boosting non-Gaussian GAMLSS. 
Our main focus will be on a negative binomial location and scale model, for which we investigate the following two aspects: 
first, the analytical approximation for $\nu_\mu$ and the base-learner based approach for $\nu_\alpha$ are assessed, where numerically determined shrunk optimal step lengths serve as the benchmark.
Like in the Gaussian simulation setting, we are not trying to beat the non-cyclical boosting algorithm with shrunk optimal step lengths, but the goal is to perform similarly well while being computationally more efficient. 
Secondly, we compare estimation results of the fixed and the adaptive step length approaches focusing on the balancedness of the overall model. \\

We consider the following model for the negative binomial response variable $y_i \sim \mathcal{NB}(\mu_i, \alpha_i)$:
\begin{align}
\eta_{\mu, i} &= \log(\mu_i) = -0.5 - 0.5 \cdot x_{1i} + 0.3 \cdot x_{2i} + 0.5 \cdot x_{4i} - 0.3 \cdot x_{5i} \nonumber \\
\eta_{\alpha, i} &= \log(\alpha_i) = 0.6 \cdot x_{2i} - 0.6 \cdot x_{3i} - 0.4 \cdot x_{5i} + 0.4 \cdot x_{6i} 
\label{daub:formula_simulation_model_negbinom}
\end{align}
Note that $x_{2}$ and $x_{5}$ are shared between both predictors and that $y_i$ has a variance of $\mu_i + \alpha_i \mu_i^2$.
We simulate $n=500$ observations of each variable, where $x_{1}, x_{2}, x_{3}$ are drawn independently from a uniform distribution on $[-1,1]$ and for $x_{4}, x_{5}, x_{6}$ independent realizations of a Bernoulli distributed random variable with $p = 0.5$ are drawn.
Except for the inclusion of binary covariates and different coefficients in order to have a setup with differing optimal step lengths, this simulation setup follows Thomas et al. (2018). 
In addition to the model in (\ref{daub:formula_simulation_model_negbinom}), we also consider settings with 10 and 150 additional non-informative covariates. \\
Compared to the Gaussian simulation, we additionally apply a non-cyclical boosting algorithm 
in which $\nu_\mu$ and $\nu_\alpha$ are both determined via line search and for the base-learner based approach with a fixed reference step length we use $\mu$ as reference parameter instead of the scale parameter.
Hence, the following step length schemes will be considered:
\begin{itemize}
\item numerically obtained shrunk optimal step lengths analogous to \cite{Zhang2022} (LS-LS):\\
$\nu_\mu$ and $\nu_\alpha$ are determined via line search on $[0,20]$ and $[0,200]$, respectively.
\item shrunk optimal step lengths using the proposed approximative analytical solution (A-LS): \\ 
$\nu_\mu$ is computed based on the analytical approximation from (\ref{daub:formula_opt_step_length_mu_neg_binom}), $\nu_\alpha$ is determined via line search on $[0,200]$.
\item new base-learner based approach with optimal reference step length (A-BL):\\
$\nu_\mu$ is computed based on the analytical approximation from (\ref{daub:formula_opt_step_length_mu_neg_binom}), $\nu_\alpha$ is computed with the base-learner based approach, see (\ref{daub:formula_BL_steplengths}).
\item new base-learner based approach with fixed reference step length (F-BL):\\
$\nu_\mu$ is defined as fixed step length of 0.1, $\nu_\alpha$ is computed with the base-learner based approach, see (\ref{daub:formula_BL_steplengths}).
\item fixed step lengths as in the initially proposal of the non-cyclical boosting algorithm (F-F): \\
$\nu_\mu$ and $\nu_\alpha$ are defined as fixed step lengths of 0.1.
\end{itemize}

\begin{figure}[b!]
\includegraphics[width=\textwidth]{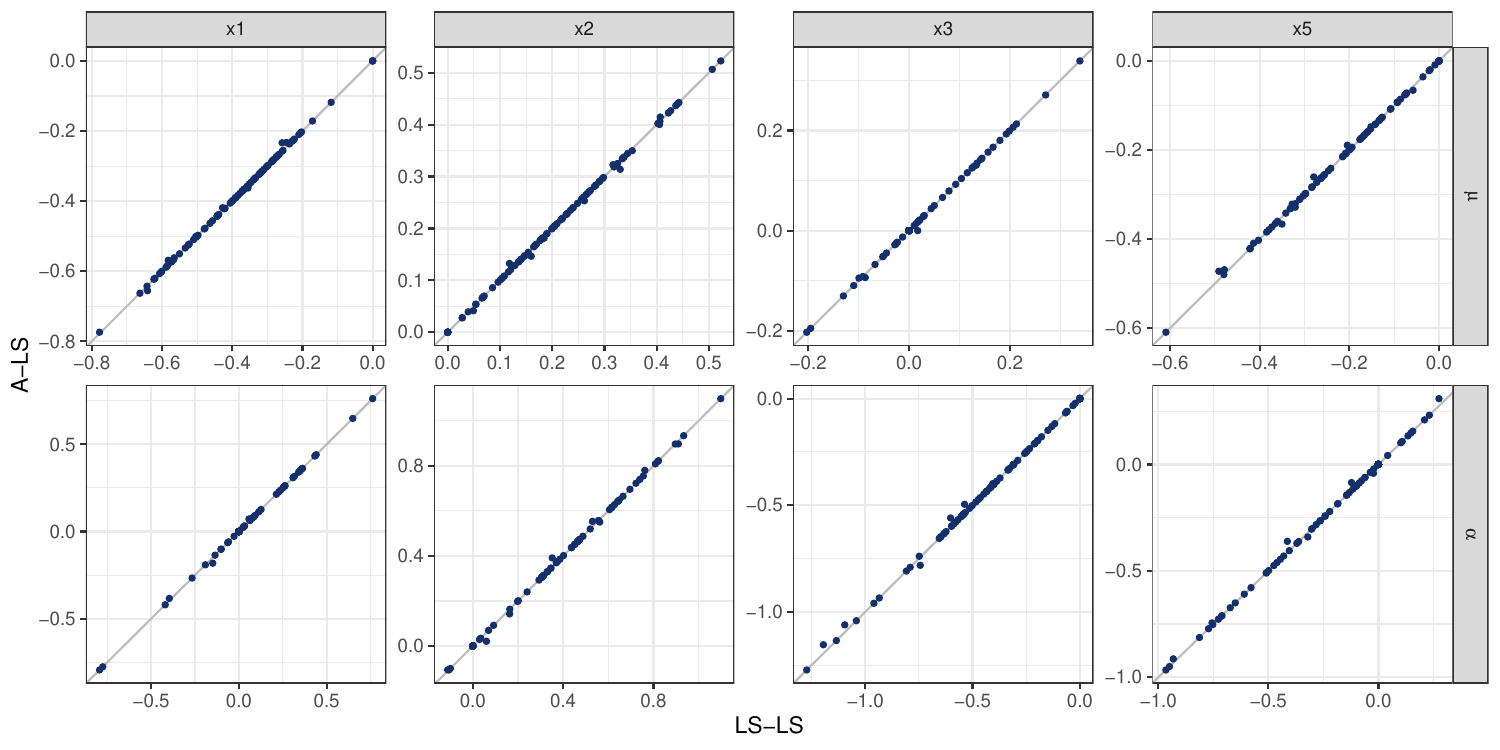}
\caption{\label{daub:fig_scatter_selCoefs_negbinom} Direct comparison of coefficient estimates of the two adaptive step length approaches LS-LS and A-LS in the negative binomial simulation setting (\ref{daub:formula_simulation_model_negbinom}) without additional non-informative covariates.}
\end{figure}

\noindent Like in the Gaussian simulation, we use a shrinkage parameter $\lambda_s$ of 0.1, determine the stopping iteration by 10-fold cross-validation and conduct $B=100$ simulation runs. \\ 

\begin{figure}[t!] 
\centering
\includegraphics[width= 0.93\textwidth]{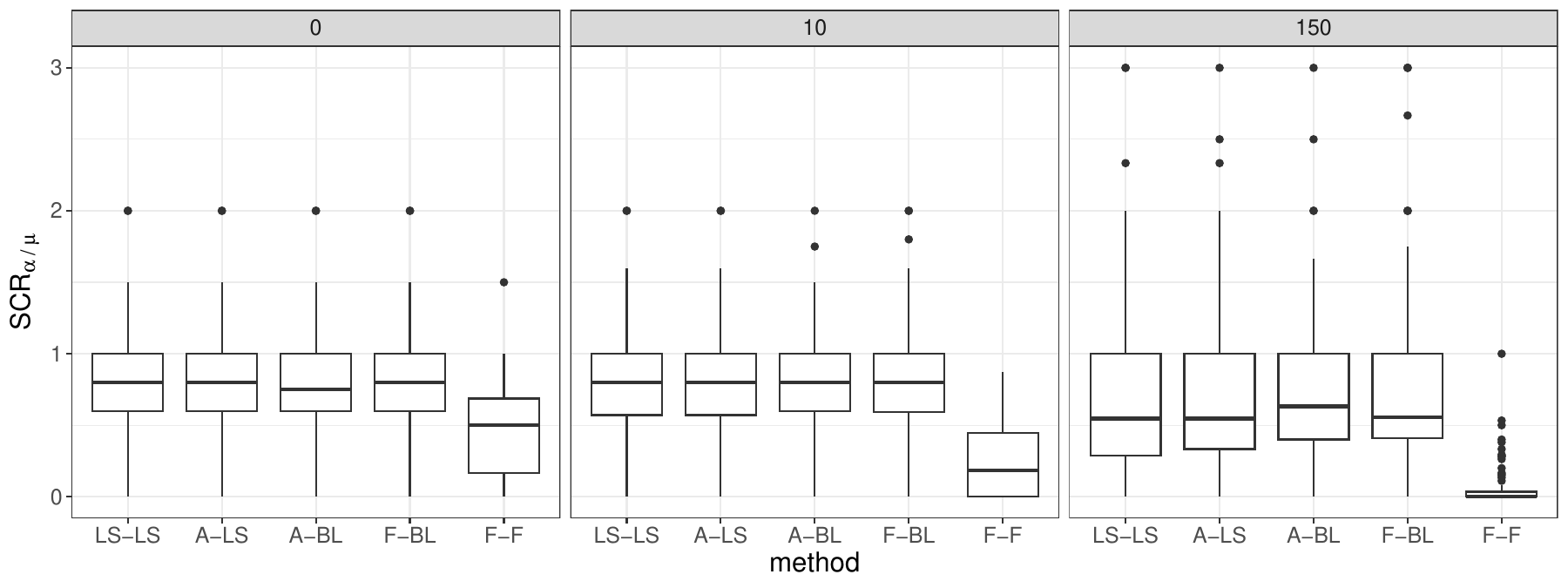}
\caption{\label{daub:fig_overview_selCoefs_raio_negbinom} Distribution of the ratio of numbers of covariates included in $\eta_\alpha$ relative to $\eta_\mu$ (SCR$_{\alpha/\mu}$) in the negative binomial simulation setting (\ref{daub:formula_simulation_model_negbinom}) for a varying number of additional non-informative covariates (columns).
15 outliers that exceed an SCR$_{\alpha/\mu}$ of 3 are regarded in the distributions but not included in the plot above for a better comprehensibility. The excluded outliers occured in 5 simulation runs with 150 additional non-informative covariates. }
\end{figure}

In order to evaluate the analytical approximation for $\nu_\mu^*$, we first compare the results of A-LS with LS\nobreakdash-LS, where we find that the resulting coefficients are almost identical over all simulation settings.
Figure~\ref{daub:fig_scatter_selCoefs_negbinom} displays the direct comparison of selected coefficients without additional non-informative covariates. 
Also with respect to the variable selection and the relative number of selected covariates for the different submodels, the two adaptive step length approaches yield very similar results (see Figure~\ref{daub:fig_overview_selCoefs_raio_negbinom} as well as Figures~\ref{daub:fig_overview_FP_negbinom} and \ref{daub:fig_overview_FN_negbinom} in Appendix~\ref{daub:appendix_simu_negbinom}).
A comparison of the base-learner based and shrunk optimal step length approach for $\nu_\alpha$ (A-BL and A-LS) shows a similar picture as in the Gaussian simulation setting
(see  Figure~\ref{daub:fig_scatter_selCoefs_ALAD_ALBL_negbinom} in Appendix~\ref{daub:appendix_simu_negbinom} for the direct comparison of selected coefficients as well as Figure~\ref{daub:fig_overview_selCoefs_raio_negbinom} and Figures~\ref{daub:fig_overview_FP_negbinom} and \ref{daub:fig_overview_FN_negbinom} in Appendix~\ref{daub:appendix_simu_negbinom} for measures of the variable selection performance and balancedness).
Using base-learner based instead of shrunk optimal step lengths thus does not affect the estimation results by a relevant degree in this setup. \\

In Gaussian models with a large variance, the size of the fitted base-learners differs substantially between the two predictors, which is also the case in the negative binomial model considered here. 
We next demonstrate that, like in the Gaussian case, this issue leads to an imbalanced overall model when using fixed step lengths and investigate the balancedness of models resulting from the adaptive step length approaches. 
The ratio of the number of covariates included in $\eta_\alpha$ relative to $\eta_\mu$ (SCR$_{\alpha/\mu}$) indicates that with a fixed step length approach $\eta_\alpha$ is underrepresentated compared to $\eta_\mu$, which becomes more pronounced the more non-informative covariates are considered (see Figure~\ref{daub:fig_overview_selCoefs_raio_negbinom}).
The adaptive step length approaches on the other hand result in a considerably more balanced overall model, where the SCR$_{\alpha/\mu}$ only exhibits a slight decline as the number of additional non-informative covariates is increased.
The imbalance in the overall model is also apparent in the distributions of the estimated coefficients (see Figure~\ref{daub:fig_overview_coeffs_negbinom} in Appendix~\ref{daub:appendix_simu_negbinom}) 
as for fixed step lengths the coefficients in $\eta_\alpha$ are considerably more shrunk towards 0 than the coefficients in $\eta_\mu$.
In accordance with that, the number of false positives in $\eta_\mu$ is considerably higher than in $\eta_\alpha$ in the fixed step length approach, while they are on a similar level for the adaptive step length approaches (see Figures~\ref{daub:fig_overview_FP_negbinom} and \ref{daub:fig_overview_FN_negbinom} in Appendix~\ref{daub:appendix_simu_negbinom}). \\

Finally, we check whether the approximations achieve the desired reduction in run time compared to numerically obtained shrunk optimal step lengths (LS-LS). 
When comparing the median run time, we find that using base-learner based step lengths with an analytically derived reference step length (A-BL) reduces the run time by about 50\% in the considered simulation setup without additional non-informative covariates (see Table~\ref{daub:table_run_time_negbinom}). 
Due to the considerably larger stopping iteration, using fixed step lengths instead of numerically obtained shrunk optimal ones takes approximately twice as long. \\ 

\begin{table}[!htb] 
      \centering
\begin{tabular}{llllll}
  \toprule[0.09 em]
 & \hspace*{0.4em} LS-LS & \hspace*{0.6em} A-LS & \hspace*{0.6em} A-BL & \hspace*{0.6em} F-BL & \hspace*{1.2em} F-F \\ 
  \midrule
1st Qu. & 0.307 (64) & 0.221 (64) & 0.125 (61) & 0.368 (166) & 0.374 (208) \\ 
  Median & 0.498 (96) & 0.358 (98) & 0.254 (99) & 0.587 (284) & 0.990 (524) \\ 
  3rd Qu. & 0.691 (134) & 0.515 (133) & 0.421 (142) & 1.023 (398) & 2.159 (1,042) \\   
   \bottomrule[0.09 em]
\end{tabular}
\caption{\label{daub:table_run_time_negbinom}
Quartiles of the run times until $m_\text{stop}$ in seconds and stopping iterations (in parenthesis) in the negative binomial simulation setting (\ref{daub:formula_simulation_model_negbinom}) without additional non-informative covariates.}
\end{table}

Both, base-learner based and shrunk optimal step lengths thus yield a considerably more balanced overall model than the fixed step length approach also in a negative binomial stimulation setting. 
Like for a Gaussian response variable, shrunk optimal and base-learner based approaches have very similar estimation results while they differ with respect to the computational efficiency.
Simulation runs with a Weibull scale and shape model show a very similar picture. 
For information on the specific simulation setting and an overview of the results, see Appendix~\ref{daub:appendix_simu_weibull}. \\

Overall, the simulation results thus confirm that the proposed base-learner based step lengths induce a natural balance between the submodels in different GAMLSS settings.
With respect to the computational efficiency, the new approach with a shrunk optimal reference step length yields the desired reduction in run time, where
for a fixed reference step length the run time is driven by the stopping iteration. 
In case of a comparatively large stopping iteration for a fixed reference step length, adjustments in the level of the reference step length can reduce stopping iteration and run time. \\

%% file: application.tex
\newpage

\section{Applications}
\label{daub:section_applications}

In the following, we will illustrate how the new step length approaches perform on two real-world data sets.
Specifically, the estimation results for a negative binomial location and scale model of the number of doctor's visits as well as for a Weibull scale and shape model of survival times will be compared with respect to, e.g., the coefficient estimates, the variable selection behavior and the predictive performance.
Based on the second data set, we moreover demonstrate the applicability to high-dimensional data.

\subsection{Modeling the number of doctor's visits}
\label{daub:subsec_Data_HealthAustralia}

First, we apply the non-cyclical boosting algorithm with different step length schemes to estimate a negative binomial location and scale model. 
Specifically, mean and overdispersion of the number of doctor's visits in Australia are modeled by different, mainly health and health insurance related characteristics. 
Covariates referring to an individual's health condition are for example the number of days with reduced activity (\textit{actdays}), whether an individual suffers from a chronic health condition (\textit{chcond1}, \textit{chcond2}) or information on recent treatments like the number of prescribed and total medications (\textit{prescrib}, \textit{medicine}). 
With respect to the health insurance, the standard health insurance with an income-based insurance levy is the baseline, whereas individuals with \textit{levyplus} are private patients with better health insurance services and individuals with \textit{freepoor} or \textit{freepera} are exempt from the levy due to for example old age or low income.
The health insurance services of individuals with \textit{freepoor} and \textit{freepera} are however similar to the standard health insurance's services. \\
The data were collected within the Australian Health Survey in 1977-1978 and comprise information on the number of doctor's visits as well as on 16 covariates for 5,190 individuals \citep{Cameron1988}.
An overview of the different variables including short explanations can be found in Appendix~\ref{daub:appendix_info_data_healthAustralia}. \\

\begin{table}[b!] 
\centering
\begin{tabular}{c|ccc|ccc}
  \toprule[0.09 em]
  & & $\eta_\mu$ & & & $\eta_\alpha$ & \\[0.3em]
 & A-LS & A-BL & F-F & A-LS & A-BL & F-F \\ 
  \midrule
(Intercept) & -2.009 & -2.062 & -2.099 & 0.696 & 0.346 & 0.257 \\ 
  sex & 0.110 & 0.122 & 0.142 & 0 & 0 & 0 \\ 
  age & 0 & 0 & 0 & 0 & 0 & 0 \\ 
  income & -0.004 & -0.007 & -0.012 & 0 & 0 & 0 \\ 
  illness & 0.137 & 0.142 & 0.148 & -0.052 & 0 & 0 \\ 
  actdays & 0.110 & 0.114 & 0.117 & 0.012 & 0.020 & 0.022 \\  
  hospadmi & 0.134 & 0.147 & 0.158 & 0.068 & 0.075 & 0.063 \\ 
  hospdays & 0.002 & 0.002 & 0.002 & 0& 0 & 0 \\ 
  medicine & 0 & 0 & -0.016 & -0.078 & -0.083 & -0.100 \\ 
  prescrib & 0.150 & 0.162 & 0.177 & -0.254 & -0.144 & -0.089 \\ 
  nondocco & 0.022 & 0.022 & 0.023 & 0 & 0& 0 \\ 
  levyplus & 0 & 0 & 0 & -0.084 & -0.046 & 0 \\ 
  freepoor & -0.201 & -0.268 & -0.347 & 0.290 & 0 & 0 \\ 
  freepera & 0.019 & 0.028 & 0.034 & 0 & 0 & 0 \\   
  hscore & 0.020 & 0.022 & 0.025 & 0 & 0 & 0 \\ 
  chcond1 & 0 & 0 & 0.008 & 0 & 0 & 0 \\ 
  chcond2 & 0 & 0 & 0 & 0 & 0 & 0 \\
  \bottomrule[0.09 em]
\end{tabular}
\caption{\label{daub:table_coefficient_estimates} Coefficient estimates at the median stopping iteration for the Australian health care data applying different step length approaches.}
\end{table}

In the following, the coefficient estimates and technical properties of non-cyclical boosting algorithms with analytically derived $\nu_\mu$ will be compared, where the step lengths for updates of $\eta_\alpha$ are either obtained by line-search (A-LS) or via the base-learner based approach (A-BL). 
Additionally, fixed step lengths of 0.1 (F-F) are included in the comparison. 
The stopping iteration is obtained as median from cross-validations on 100 different randomly drawn folds and we choose a shrinkage factor of $\lambda_s = 0.1$. \\
With respect to the median stopping iteration, the two adaptive step length approaches yield comparable results, where however A-BL runs a little longer due to slightly more moderate update sizes. 
When using fixed step lengths, the algorithm is stopped in median after about 7.5 times as many iterations as for A-BL. 
In accordance with that, the run time of F-F until the median $m_\text{stop}$ is about 7 times as high, while A-BL reduces the run time of A-LS until the median $m_\text{stop}$ by about 15\% despite the later stopping iteration. 
The number of iterations until stopping and the run times until $m_\text{stop}$ can be found in Appendix~\ref{daub:appendix_healthAustralia_technical_properties}. \\

Table~\ref{daub:table_coefficient_estimates} displays the coefficient estimates at the median $m_\text{stop}$ for different step length approaches. 
Regarding the effect of different covariates on the distribution of the number of doctor's visits, we for example find that a larger number of prescribed medications (\textit{prescrib}) goes along with a higher expected number of doctor's visits and a lower overdispersion. 
As the negative effect of \textit{prescrib} on the overdispersion partly compensates the positive effect on the mean, \textit{prescrib} has a small positive effect on the variance of the number of doctor's visits. 
Conversely, an increase in the number of days of reduced activity (\textit{actdays}) goes along with an increase in both the mean and the overdispersion of the number of doctor's visits and thus implies a larger increase in the dependent variable's variance. \\ 

\begin{figure}[t!]
\includegraphics[width=\textwidth]{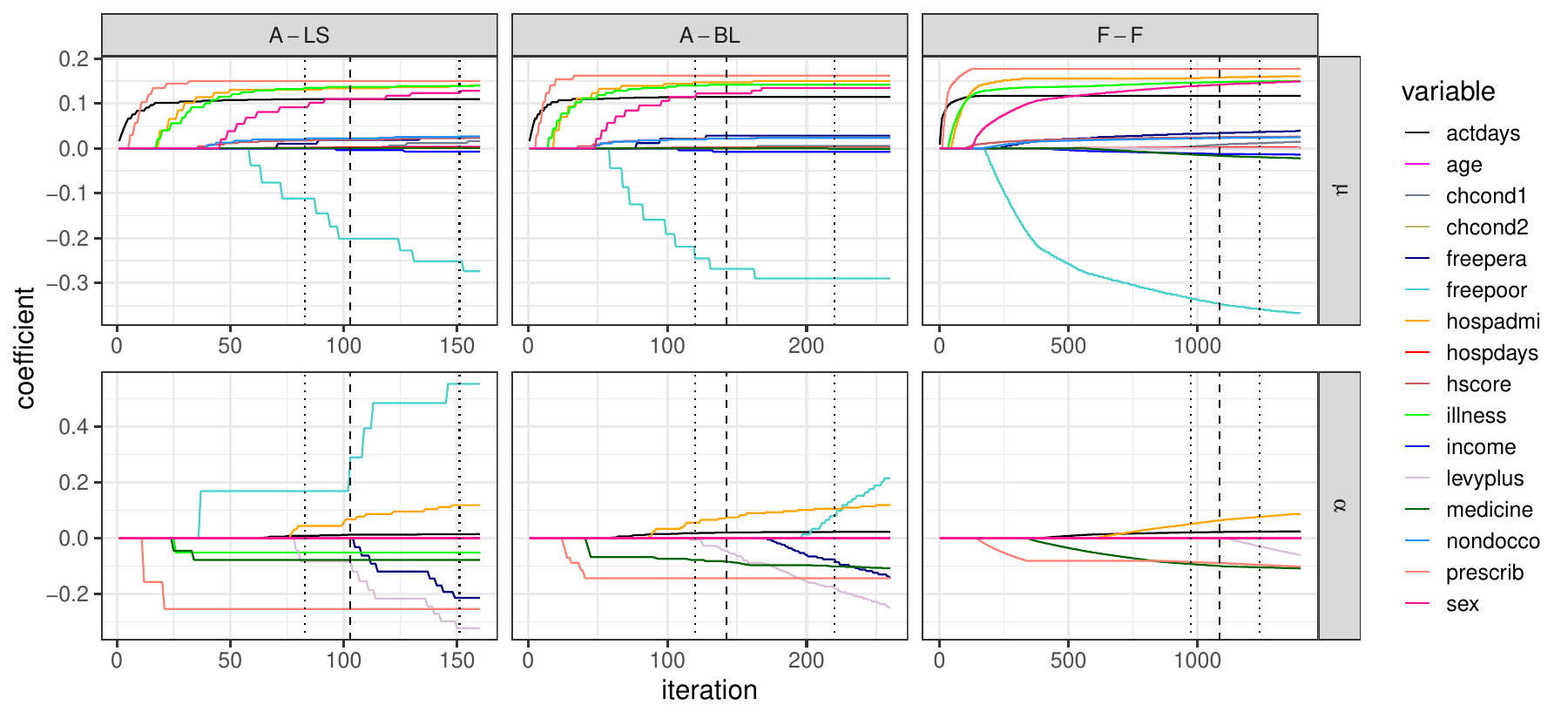}
\caption{\label{daub:fig_healthAustralia_coeff_paths} Coefficient paths for the Australian health care data using different step length approaches (columns). 
The vertical lines represent the median stopping iteration (dashed) as well as the 1st and 3rd quartile (dotted) of the stopping iteration obtained via cross-validations on 100 randomly drawn folds.
Please be aware of the differences in the x-axis scaling.} 
\end{figure}

In order to compare the different step length approaches in more detail, in addition to the coefficient estimates at the median $m_\text{stop}$ we also take the entire coefficient paths into consideration, which are depicted in Figure~\ref{daub:fig_healthAustralia_coeff_paths}. 
Comparing the two adaptive step length approaches, we find that overall the paths of A-LS and A-BL exhibit a similar behavior where the paths of A-BL grow slightly slower and differences in the paths of $\eta_\alpha$ are apparent for certain covariates. 
These are impacted by the following aspects. 
On the one hand, correlated covariates in combination with differences in the update size during the first iterations can result in differences in the estimated effects corresponding to the respective covariates. 
In this application, \textit{prescrib} and \textit{medicine} are highly correlated (see Appendix~\ref{daub:appendix_healthAustralia_corr} for the correlation plot). 
That results in the first large steps of \textit{prescrib} in A-LS leading to \textit{medicine} being updated less later on, while \textit{prescrib} and \textit{medicine} have a more similar impact on $\eta_\alpha$ for A-BL step lengths, which yield more moderate update sizes in the beginning. 
On the other hand, update sizes can be overproportionally large for certain covariates when using shrunk optimal step lengths, which is not captured by the base-learner based approach.
As this is the case for \textit{freepoor} in this application, its path in $\eta_\alpha$ starts earlier and is steeper for A-LS than for A-BL.
The differences in update size originate from different selection criteria for the best-fitting covariate of a submodel (sum of squared residuals, see step 3 of Algorithm~\ref{daub:algo_non-cyclical_boosting_GAMLSS}) and the predictor that is updated (reduction of the loss function, see step 6 of Algorithm~\ref{daub:algo_non-cyclical_boosting_GAMLSS}) in the non-cyclical boosting algorithm. \\ 

From Figure~\ref{daub:fig_healthAustralia_coeff_paths} and Table~\ref{daub:table_coefficient_estimates} it is moreover evident that the relative importance of the two submodels differs between adaptive and fixed step length approaches.
While the coefficient paths of $\eta_\mu$ for the fixed step length approach grow with a moderately lower speed compared to the adaptive step length approaches (by a factor roughly between 2.5 and 4), the paths of $\eta_\alpha$ grow considerably slower and start later for a boosting algorithm with fixed compared to adaptive step lengths (by a factor roughly between 10 and 15). 
With fixed step lengths, updates of $\eta_\mu$ are thus favored more strongly than for the adaptive step length approaches, which indicates that the fixed step length approach exhibits the imbalancedness issue outlined in section~\ref{daub:section_nonclyclical_boosting_with_adaptive_steplengths}.
In accordance with that, the fixed step length approach selects two more covariates for $\eta_\mu$ at the median $m_\text{stop}$, while fewer covariates are included in $\eta_\alpha$ (see Tables~\ref{daub:table_coefficient_estimates} and \ref{daub:table_sel_coeffs_HealthAustralia}). \\

\begin{table}[t!] 
\centering
\footnotesize
\begin{tabular}{c|ccc|ccc}
  \toprule[0.09 em]
  & & $\eta_\mu$ & & & $\eta_\alpha$ & \\[0.3em]
 & A-LS & A-BL & F-F & A-LS & A-BL & F-F \\ 
  \midrule
1st Qu. &  10 &  11 &  13 &   7 &   5 &   4 \\ 
Median &  11 &  11 &  13 &   7 &   5 &   4 \\ 
  3rd Qu. &  13 &  13 &  13 &   9 &   7 &   5 \\   
  \bottomrule[0.09 em]
\end{tabular}
\caption{\label{daub:table_sel_coeffs_HealthAustralia} Number of selected covariates at different quartiles of the stopping iterations (rows) for the Australian health care data and for the different predictors as well as different step length approaches (columns).}
\end{table}

Overall, these results confirm the findings from the simulation study that adaptive step length approaches ensure a more balanced overall model compared to the fixed step length approach, where the results from A\nobreakdash-LS and A-BL are similar but do not fully coincide here.
Numerically obtained shrunk optimal step lengths for both $\mu$ and $\alpha$ as well as base-learner based step lengths with a fixed reference step length, which we did not discuss in detail here, yield very similar results as A-LS and A-BL step lengths, respectively (see Appendix \ref{daub:appendix_Health_Australia_LS-LS_F-BL}).

\subsection{Predicting the survival time for diffuse large-B-cell lymphoma patients}

As a second real-world data set we consider the diffuse large-B-cell lymphoma (DLBCL) data collected by \cite{Rosenwald2002}, which contain the survival times of 224 patients with diffuse large-B-cell lymphoma as well 7,399 microarray gene expression measurements after receiving chemotherapy. 
In the following analysis, only a subpopulation of 127 patients who died during the follow-up is considered. 
The median follow-up time for survivors was 7.3 years and 2.8 years for all patients. \\

The focus in this high-dimensional application will be on the variable selection behavior, especially with respect to the balancedness of the overall model.
Moreover, the step length approaches will be compared with regard to their predictive performance, where shrunk optimal step lengths (LS-LS, A-LS), base-learner based step lengths with a shrunk optimal and fixed reference step length (A-BL, F-BL) as well as a fixed step lengths of 0.1 (F-F) will be considered.
The step length approaches are specified the same way as in the beginning of section~\ref{daub:subsec_Data_HealthAustralia}.
The stopping iteration is determined via 10-fold cross-validation and a shrinkage factor of $\lambda_s=0.1$ is considered. \\

\begin{table}[b!]
\centering
\begin{tabular}{c|ccccc}
  \toprule[0.09 em]
 & LS-LS & A-LS & A-BL & F-BL & F-F \\
\midrule
	sel. cov. $\eta_\lambda$ &  10 &  10 &  10 &   8 &  12 \\ 
   sel. cov. $\eta_k$ &   6 &   6 &   7 &   4 &   4 \\ 
   \midrule
SCR$_{\lambda/k}$ & 1.67 & 1.67 & 1.43 & 2.00 & 3.00 \\    
  \bottomrule[0.09 em]
\end{tabular}
\caption{\label{daub:table_sel_coeffs_DLBCL} Number of selected covariates for the different predictors including their ratio (rows) as well as different step length approaches (columns) for the DLBCL data.}
\end{table}

Similar to the previously considered Australian health care data, all approaches select more covariates for $\eta_\lambda$ than for $\eta_k$, which is more pronounced in the fixed step length approach (see Table~\ref{daub:table_sel_coeffs_DLBCL}). 
The higher relative importance of $\eta_\lambda$ in the overall model when using fixed step lengths translates to an SCR$_{\lambda/k}$ of about twice the size of the adaptive step length approaches, while base-learner based and shrunk optimal step lengths yield similar SCR$_{\lambda/k}$ levels. 
Note that even though the differences in the optimal step lengths are not particularly large here, they have a relevant impact on the overall model's balancedness since the algorithm is stopped after comparatively few iterations due to the high-dimensional setting. \\
With respect to the absolute numbers of selected covariates it is apparent that the F-BL approach includes fewer covariates in both submodels than the other adaptive step length approaches, which could be explained by the fixed step length of 0.1 being too aggressive for the data at hand.
Note moreover that base-learner based and shrunk optimal step length approaches do not always select the same covariates, which is not unexpected for data with many and correlated covariates. 
For the specific coefficient estimates, see Tables~\ref{daub:table_coeffs_lambda_DLBCL} and \ref{daub:table_coeffs_k_DLBCL} in Appendix \ref{daub:appendix_DLBCL}. 
Even though not all selected covariates coincide, 
the predictive performance is similar. \\ 

\begin{figure}[!htb]
\centering 
\includegraphics[width=0.56\textwidth]{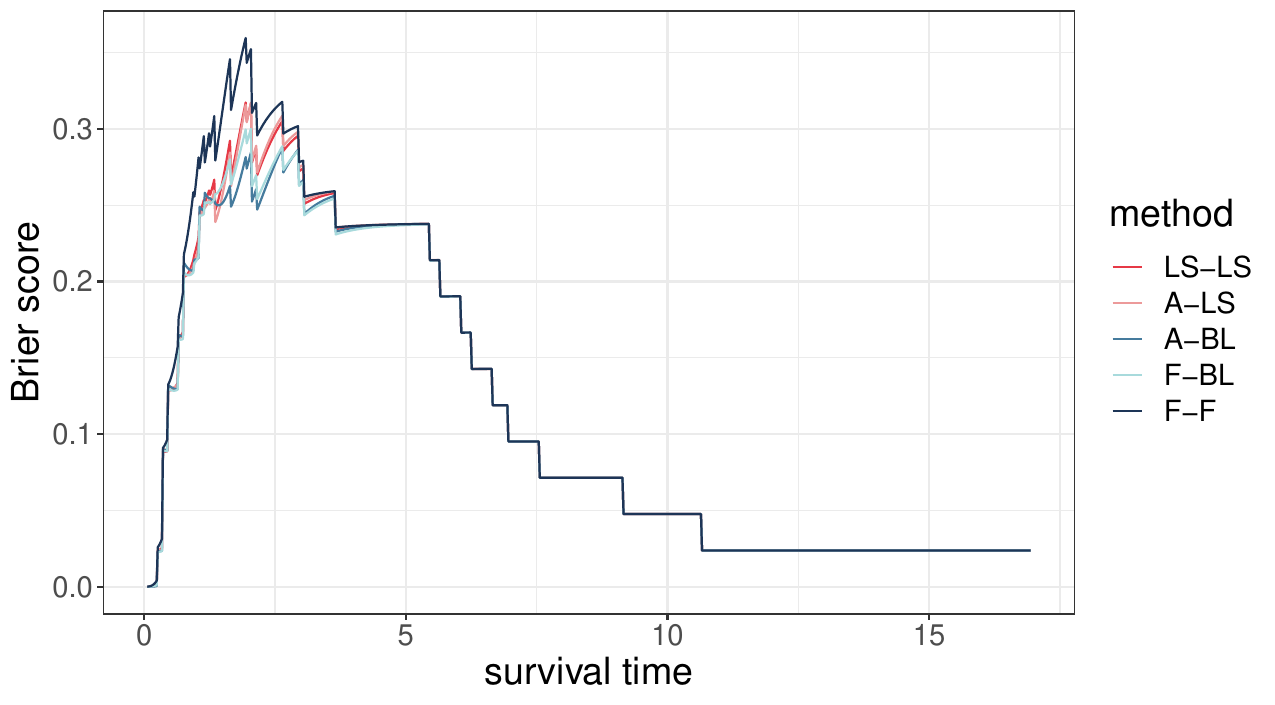}
\hfill
\includegraphics[width=0.41\textwidth]{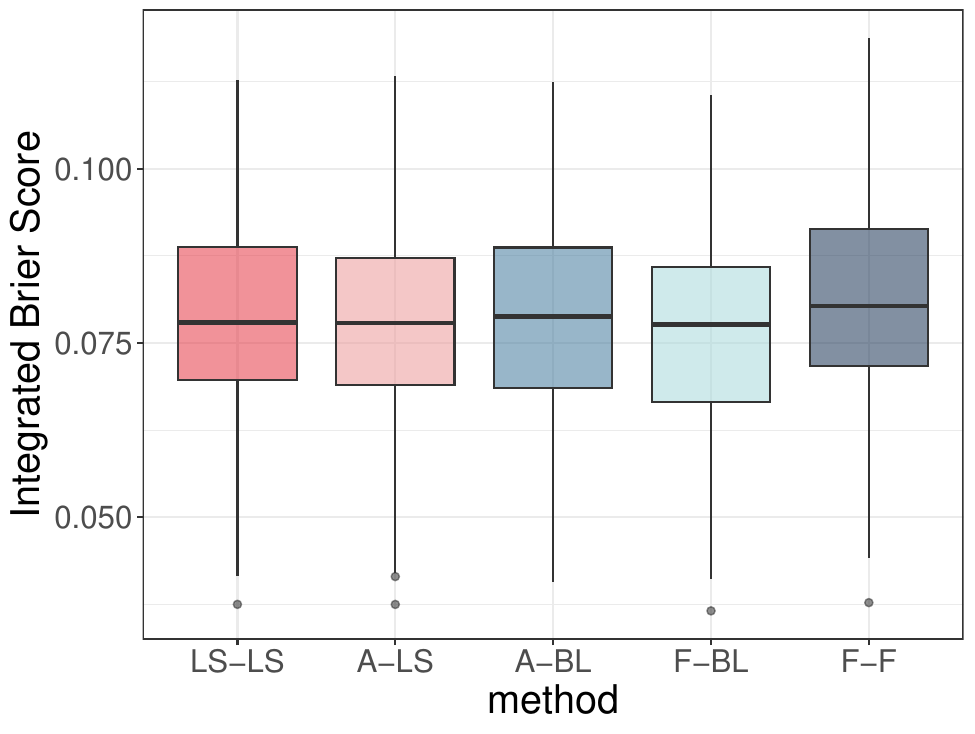}
\caption{\label{daub:fig_DLBCL_Brier_scores} Brier scores for a single split into training and validation set (left) as well as the distributions of the Integrated Brier Score for 100 random splits (right) for the DLBCL data using different step length approaches.} 
\end{figure}

Figure~\ref{daub:fig_DLBCL_Brier_scores} depicts the Brier scores \citep{Brier1950} for a single random split into training and validation set (left) as well as the distribution of the Integrated Brier Score \citep{Graf1999} for 100 splits (right), where following \cite{Rosenwald2002} the validation set comprises about one third of the data.
From Figure~\ref{daub:fig_DLBCL_Brier_scores} is evident that all step length approaches yield similar results with respect to the Integrated Brier Score, where the fixed step length approach performs slightly worse than the adaptive step length approaches and little differences in the distributions among the adaptive step length schemes are apparent. 
For a single run, the Brier scores show that the main difference between the step length approaches relates to predictions for survival times between 0.5 and 3.5 years and that the two shrunk optimal step length approaches as well as the base-learner based approaches seem to yield particularly similar Brier scores, respectively.
With the Brier score of a trivial model being 0.25 it is moreover evident that all models have difficulties with predictions in this range. \\

%% file: conclusion.tex
\section{Conclusion and Discussion}
\label{daub:section_conclusion}

Shrunk optimal step lengths are known to address the issue of imbalanced submodel updates when fitting a GAMLSS by means of a non-cyclical boosting algorithm. 
As alternative to having to solve the respective optimization problem for every potential update either analytically or numerically, we introduced a new step length approach that determines the adaptive step lengths from algorithm intrinsic information. 
As it uses the inverse ratio of base-learner norms as step length, the new adaptive step length approach ensures naturally balanced submodel updates by fixing all potential updates to the same size.
In addition to a Gaussian location and scale model, for which shrunk optimal step lengths have been introduced by \cite{Zhang2022}, we implemented the different adaptive step length approaches for negative binomial and Weibull distributed response variables.
For that, an approximate optimal step length for the location parameter $\mu$ in a negative binomial location and scale model as well as for the scale parameter $\lambda$ in a Weibull scale and shape model was derived. \\

By means of a simulation study as well as with real-world data sets we could show that the proposed base-learner based step length approach with either a shrunk optimal or a small fixed reference step length solves the imbalancedness problem that can arise when using fixed step lengths. 
The simulation results moreover indicate that base-learner based approaches yield similar results as the shrunk optimal step length approach with respect to the coefficient estimates as well as with respect to the variable selection while reducing the run time compared to numerically obtained shrunk optimal step lengths. 
When the data is correlated, differences between the adaptive step length approaches can occur as became evident in the applications.
Especially in the high-dimensional setting, the coefficient estimates resulting from the different adaptive step length approaches do not always coincide and different variables can be selected.
With respect to the analytically derived approximate optimal step lengths, 
the estimation results were almost identical compared to numerically determined optimal step lengths. \\

A limitation of the proposed balanced boosting approach for GAMLSS is that it is currently only available for linear base-learners. 
Future plans include extending the approach to other effect types, e.g., non-linear or spatial effects as well as to further response variable distributions.
Along with that, both adaptive step length approaches are planned to be included into the R add-on package gamboostLSS \citep{Hofner2016} in order to facilitate access. \\ 
Another limitation is that the algorithm tends to include too many covariates in some situations, which is a common problem of statistical boosting algorithms and could also be observed for shrunk optimal and fixed step lengths.
Since by using adaptive step lengths all submodels exhibit a similar degree of convergence, this issue could be addressed by a more robust stopping iteration \citep[see e.g.][]{Ellenbach2021}.
A different approach to obtain a sparser model, which effectively deselects covariates of minor importance, was recently proposed by \cite{Stroemer2022}.
In particular the effect of the proposed deselection routine on the overall model's balancedness would be interesting to investigate in future research.\\

In summary, using base-learner based step lengths has the following advantages: 
most importantly, it ensures a natural balance between the different submodels, which is not the case for fixed step lengths that are typically used by default.
In addition, the proposed step length approach is computationally more efficient than using shrunk optimal step lengths and straight-forward to apply in any type of boosting algorithm with multiple predictors. 
We are therefore optimistic that using base-learner based step lengths can help practioners to address update selection issues in a broad range of boosting algorithms when multiple submodels have to be fitted like distributional copula regression \citep{Hans2022} or multivariate distributional regression \citep{Stroemer2023}. 

%% file: appendix.tex
\noindent {\Huge \textbf{Appendices} }

\section{Derivation of the approximated optimal step lengths}

\subsection{Derivation of $\boldsymbol{\nu_\mu^*}$ for a negative binomial response variable}
\label{daub:appendix_opt_steplength_neg_binom}

Starting from the probability mass function in (\ref{daub:formula_pmf_neg_binom}), 
we have a log-likelihood function of the negative binomial location and scale model of \citep{Hilbe2011}
\begin{align*}
\ell &(\boldsymbol{\mu}, \boldsymbol{\alpha}; \boldsymbol{y}) 
= \sum_{i=1}^n y_i \ln \left( \frac{\alpha_i \mu_i}{1+\alpha_i \mu_i}\right) - \frac{1}{\alpha_i} \ln(1+\alpha_i \mu_i) + \ln\Gamma\left(y_i+\frac{1}{\alpha_i}\right) - \ln\Gamma(y_i+1) - \ln\Gamma\left(\frac{1}{\alpha_i}\right) \, ,
\end{align*}
where $\Gamma(\cdot)$ denotes the gamma function. \\
In order to determine the optimal step length $\nu_\mu^{*}$ in iteration $m$, the following optimization problem with the negative log-likelihood $-\ell(\cdot)$ as the loss function has to be solved
\begin{align}
\nu^{*[m]}_\mu = \underset{\nu_\mu}{\text{arg min}} \, -\ell &\left(\exp\left(\boldsymbol{\eta}_\mu^{[m-1]} + \nu_\mu \, \boldsymbol{h}^{[m]}_{j_\mu^*, \mu} \right), \boldsymbol{\alpha}^{[m-1]}; \boldsymbol{y} \right) \, ,
\label{daub:formula_appendix_opt_problem_nu_mu_neg_binom}
\end{align}
where $\exp(\cdot)$ is applied element-wise. \\
For the first order condition holds
\begin{align*}
0 = - \frac{\partial \ell}{\partial \nu_\mu} \Bigg\vert_{\nu_\mu=\nu_\mu^{*[m]}}
&= - \sum_{i=1}^n \left( \frac{\partial \ell}{\partial \mu^{[m]}_i}\right) \left(\frac{\partial \mu^{[m]}_i}{\partial \eta_{\mu,i}^{[m]}}\right) \left(\frac{\partial \eta_{\mu,i}^{[m]}}{\partial \nu_\mu}\right)
\end{align*}
We obtain for the first partial derivative $\frac{\partial \ell}{\partial \mu^{[m]}_i}$:
\begin{align*}
\frac{\partial \ell}{\partial \mu^{[m]}_i}
= \frac{y_i}{\mu^{[m]}_i \left(1 + \alpha^{[m-1]}_i \mu^{[m]}_i\right)} - \frac{1}{1 + \alpha^{[m-1]}_i \mu^{[m]}_i} 
= \frac{y_i - \mu^{[m]}_i}{\mu^{[m]}_i \left(1 + \alpha^{[m-1]}_i \mu^{[m]}_i\right)} 
\end{align*}
With $\mu^{[m]}_i = \exp\left(\eta^{[m]}_{\mu,i}\right)$ and $\eta^{[m]}_{\mu,i} = \eta_{\mu,i}^{[m-1]} + \nu_\mu \, h_{\mu,i}^{*[m]}$, where we define  $\boldsymbol{h}^{[m]}_{j_\mu^*, \mu} = \left(h_{\mu,i}^{*[m]}\right)_{1 \leq i \leq n}$ for notational convenience, we furthermore have
\begin{align*}
\frac{\partial \mu^{[m]}_i}{\partial \eta^{[m]}_{\mu,i}} = \exp\left(\eta^{[m]}_{\mu,i}\right) = \mu^{[m]}_i \quad \text{and} \quad \frac{\partial \eta^{[m]}_{\mu,i}}{\partial \nu_\mu} = h_{\mu,i}^{*[m]}.
\end{align*}
The first order condition of (\ref{daub:formula_appendix_opt_problem_nu_mu_neg_binom}) thus is
\begin{align*}
0 = - \frac{\partial \ell}{\partial \nu_\mu} \Bigg\vert_{\nu_\mu=\nu_\mu^{*[m]}}
&= - \sum_{i=1}^n \frac{y_i - \mu^{[m]}_i}{\mu^{[m]}_i \, \left(1 + \alpha^{[m-1]}_i \mu^{[m]}_i\right)} \cdot \mu^{[m]}_i \cdot h_{\mu,i}^{*[m]} \\
&= - \sum_{i=1}^n \frac{h_{\mu,i}^{*[m]} \left(y_i - \exp\left(\eta_{\mu,i}^{[m-1]} + \nu_\mu^{*[m]} h_{\mu,i}^{*[m]}\right)\right)}{1 + \alpha_i^{[m-1]} \exp\left(\eta_{\mu,i}^{[m-1]} + \nu_\mu^{*[m]} h_{\mu,i}^{*[m]}\right)} \, .
\end{align*}
Approximating $\nu_\mu^{*[m]}$ in the denominator by $\nu_\mu^{[\text{prev}, j^*]}$ and using a Taylor approximation of degree 1 around $\nu_\mu^{[\text{prev}, j^*]}$ for $\exp\left(\eta_{\mu,i}^{[m-1]} + \nu_\mu^{*[m]} \, h_{\mu,i}^{*[m]}\right)$ in the numerator, i.e.,
\begin{align*}
\exp\Big(\eta_{\mu,i}^{[m-1]} + \nu_\mu^{*[m]} \, h_{\mu,i}^{*[m]}\Big) 
\approx \exp\left(\eta_{\mu,i}^{[m-1]}\, + \, \nu_\mu^{[\text{prev}, j]} \, h_{\mu,i}^{*[m]}\right) \, \cdot
\left[ 1 + h_{\mu,i}^{*[m]} \, \left(\nu_\mu^{*[m]} - \nu_\mu^{[\text{prev}, j]}\right) \right] ,
\end{align*}
we obtain 
\begin{align*}
0 &= - \sum_{i=1}^n \frac{h_{\mu,i}^{*[m]} \left(y_i - \exp\left(\eta_{\mu,i}^{[m-1]} + \nu_\mu^{*[m]} h_{\mu,i}^{*[m]}\right)\right)}{1 + \alpha_i^{[m-1]} \exp\left(\eta_{\mu,i}^{[m-1]} + \nu_\mu^{*[m]} h_{\mu,i}^{*[m]}\right)} \\
&\approx - \sum_{i=1}^n \frac{h_{\mu,i}^{*[m]} \left(y_i - \left(\exp\left(\eta_{\mu,i}^{[m-1]}\, + \, \nu_\mu^{[\text{prev}, j]} \, h_{\mu,i}^{*[m]}\right) \, \cdot
\left[ 1 + h_{\mu,i}^{*[m]} \, \left(\nu_\mu^{*[m]} - \nu_\mu^{[\text{prev}, j]}\right) \right]\right)\right)}{1 + \alpha_i^{[m-1]} \exp\left(\eta_{\mu,i}^{[m-1]} + \nu_\mu^{[\text{prev}, j^*]} h_{\mu,i}^{*[m]}\right)} \\
&= - \sum_{i=1}^n \frac{h_{\mu,i}^{*[m]} \left(y_i - \left(\exp\left(\eta_{\mu,i}^{[m-1]}\, + \, \nu_\mu^{[\text{prev}, j]} \, h_{\mu,i}^{*[m]}\right) \, \cdot
\left[ 1 - h_{\mu,i}^{*[m]} \, \nu_\mu^{[\text{prev}, j]} \right]\right)\right)}{1 + \alpha_i^{[m-1]} \exp\left(\eta_{\mu,i}^{[m-1]} + \nu_\mu^{[\text{prev}, j^*]} h_{\mu,i}^{*[m]}\right)} \\
&\hspace*{1.5em} + \nu_\mu^{*[m]} \sum_{i=1}^n \frac{{h_{\mu,i}^{*[m]}}^2 \exp\left(\eta_{\mu,i}^{[m-1]}\, + \, \nu_\mu^{[\text{prev}, j]} \, h_{\mu,i}^{*[m]}\right)}{1 + \alpha_i^{[m-1]} \exp\left(\eta_{\mu,i}^{[m-1]} + \nu_\mu^{[\text{prev}, j^*]} h_{\mu,i}^{*[m]}\right)} \, .
\end{align*}
For the optimal step length we therefore obtain:
\begin{align*}
\nu_\mu^{*[m]} \approx 
\frac{\sum_{i=1}^n \frac{h_{\mu,i}^{*[m]} \left(y_i - \left(\exp\left(\eta_{\mu,i}^{[m-1]}\, + \, \nu_\mu^{[\text{prev}, j]} \, h_{\mu,i}^{*[m]}\right) \, \cdot
\left[ 1 - h_{\mu,i}^{*[m]} \, \nu_\mu^{[\text{prev}, j]} \right]\right)\right)}{1 + \alpha_i^{[m-1]} \exp\left(\eta_{\mu,i}^{[m-1]} + \nu_\mu^{[\text{prev}, j^*]} h_{\mu,i}^{*[m]}\right)}}
{\sum_{i=1}^n \frac{{h_{\mu,i}^{*[m]}}^2 \exp\left(\eta_{\mu,i}^{[m-1]}\, + \, \nu_\mu^{[\text{prev}, j]} \, h_{\mu,i}^{*[m]}\right)}{1 + \alpha_i^{[m-1]} \exp\left(\eta_{\mu,i}^{[m-1]} + \nu_\mu^{[\text{prev}, j^*]} h_{\mu,i}^{*[m]}\right)}}
\end{align*}

\subsection{Derivation of $\boldsymbol{\nu_\lambda^*}$ for a Weibull response variable}
\label{daub:appendix_opt_steplength_weibull}

With the Weibull density function 
\begin{align*}
f(y_i;\lambda_i, k_i) 
= \frac{k_i}{\lambda_i} \, \left(\frac{y_i}{\lambda_i}\right)^{k_i-1} \, \exp\left(-\left(\frac{y_i}{\lambda_i}\right)^{k_i} \right)
\end{align*}
we have a log-likelihood function of the Weibull scale and shape model of
\begin{align*}
\ell &(\boldsymbol{\lambda}, \boldsymbol{k}; \boldsymbol{y}) 
= \sum_{i=1}^n \ln(k_i) - k_i \ln(\lambda_i) + (k_i-1) \ln(y_i) - \left(\frac{y_i}{\lambda} \right)^{k_i} \, .
\end{align*}
In order to determine the optimal step length $\nu_\lambda^{*}$ in iteration $m$, the following optimization problem with the negative log-likelihood $-\ell(\cdot)$ as the loss function has to be solved
\begin{align}
\nu^{*[m]}_\lambda = \underset{\nu_\lambda}{\text{arg min}} \, -\ell &\left(\exp\left(\boldsymbol{\eta}_\lambda^{[m-1]} + \nu_\lambda \, \boldsymbol{h}^{[m]}_{j_\lambda^*, \lambda} \right), \boldsymbol{k}^{[m-1]}; \boldsymbol{y} \right) \, ,
\label{daub:formula_appendix_opt_problem_nu_lambda_weibull}
\end{align}
where $\exp(\cdot)$ is applied element-wise. \\
For the first order condition holds
\begin{align*}
0 = - \frac{\partial \ell}{\partial \nu_\lambda} \Bigg\vert_{\nu_\lambda=\nu_\lambda^{*[m]}}
&= - \sum_{i=1}^n \left( \frac{\partial \ell}{\partial \lambda^{[m]}_i}\right) \left(\frac{\partial \lambda^{[m]}_i}{\partial \eta_{\lambda,i}^{[m]}}\right) \left(\frac{\partial \eta_{\lambda,i}^{[m]}}{\partial \nu_\lambda}\right)
\end{align*}
We obtain for the first partial derivative $\frac{\partial \ell}{\partial \lambda^{[m]}_i}$:
\begin{align*}
\frac{\partial \ell}{\partial \lambda^{[m]}_i}
= \frac{k^{[m-1]}_i}{\lambda^{[m]}_i} \left[ \left(\frac{y_i}{\lambda^{[m]}_i}\right)^{k^{[m-1]}_i} - 1 \right]
\end{align*}
With $\lambda^{[m]}_i = \exp\left(\eta^{[m]}_{\lambda,i}\right)$ and $\eta^{[m]}_{\lambda,i} = \eta_{\lambda,i}^{[m-1]} + \nu_\lambda \, h_{\lambda,i}^{*[m]}$, where we define  $\boldsymbol{h}^{[m]}_{j_\lambda^*, \lambda} = \left(h_{\lambda,i}^{*[m]}\right)_{1 \leq i \leq n}$ for notational convenience, we furthermore have
\begin{align*}
\frac{\partial \lambda^{[m]}_i}{\partial \eta^{[m]}_{\lambda,i}} = \exp\left(\eta^{[m]}_{\lambda,i}\right) = \lambda^{[m]}_i \quad \text{and} \quad \frac{\partial \eta^{[m]}_{\lambda,i}}{\partial \nu_\lambda} = h_{\lambda,i}^{*[m]}.
\end{align*}
The first order condition of (\ref{daub:formula_appendix_opt_problem_nu_lambda_weibull}) thus is
\begin{align*}
0 = - \frac{\partial \ell}{\partial \nu_\lambda} \Bigg\vert_{\nu_\lambda=\nu_\lambda^{*[m]}}
&= - \sum_{i=1}^n \frac{k^{[m-1]}_i}{\lambda^{[m]}_i} \left[ \left(\frac{y_i}{\lambda^{[m]}_i}\right)^{k^{[m-1]}_i} - 1 \right] \cdot \lambda^{[m]}_i \cdot h_{\lambda,i}^{*[m]} \\
&= \sum_{i=1}^n  \, h_{\lambda,i}^{*[m]} \, k^{[m-1]}_i
- \sum_{i=1}^n  \,  h_{\lambda,i}^{*[m]} \, k^{[m-1]}_i \, y_i^{k^{[m-1]}_i} \, 
\exp\left(-k^{[m-1]}_i \, \eta_{\lambda,i}^{[m-1]} - \nu_\lambda^{*[m]} \, h_{\lambda,i}^{*[m]} \, k^{[m-1]}_i \right) .
\end{align*}
Approximating $\exp\left(-k^{[m-1]}_i \, \eta_{\lambda,i}^{[m-1]} - \nu_\lambda^{*[m]} \, h_{\lambda,i}^{*[m]} \, k^{[m-1]}_i \right)$ via a Taylor polynomial of degree 1 around $\nu_\lambda^{[\text{prev}, j^*]}$, i.e., 
\begin{align*}
&\exp\left(-k^{[m-1]}_i \, \eta_{\lambda,i}^{[m-1]} - \nu_\lambda^{*[m]} \, h_{\lambda,i}^{*[m]} \, k^{[m-1]}_i \right) \\
&\hspace*{4em} \approx 
\exp\left(\eta_{\lambda,i}^{[m-1]} \, + \, \nu_\lambda^{[\text{prev}, j]} \, h_{\lambda,i}^{*[m]}\right)^{-k^{[m-1]}_i}
\left[ 1 + k^{[m-1]}_i \, h_{\lambda,i}^{*[m]} \, \nu_\lambda^{[\text{prev}, j]}  - \nu_\lambda^{*[m]} \, k^{[m-1]}_i \, h_{\lambda,i}^{*[m]} \right]
\end{align*}
we have
\begin{align*}
0 &= \sum_{i=1}^n  \, h_{\lambda,i}^{*[m]} \, k^{[m-1]}_i
- \sum_{i=1}^n  \,  h_{\lambda,i}^{*[m]} \, k^{[m-1]}_i \, y_i^{k^{[m-1]}_i} \, 
\exp\left(-k^{[m-1]}_i \, \eta_{\lambda,i}^{[m-1]} - \nu_\lambda^{*[m]} \, h_{\lambda,i}^{*[m]} \, k^{[m-1]}_i \right) \\
&\approx \sum_{i=1}^n  \, h_{\lambda,i}^{*[m]} \, k^{[m-1]}_i \\
&\hspace{1em} - \sum_{i=1}^n  \,  h_{\lambda,i}^{*[m]} \, k^{[m-1]}_i \, y_i^{k^{[m-1]}_i} \, 
\exp\left(\eta_{\lambda,i}^{[m-1]} \, + \, \nu_\lambda^{[\text{prev}, j]} \, h_{\lambda,i}^{*[m]}\right)^{-k^{[m-1]}_i}
\left[ 1 + k^{[m-1]}_i \, h_{\lambda,i}^{*[m]} \, \nu_\lambda^{[\text{prev}, j]}  - \nu_\lambda^{*[m]} \, k^{[m-1]}_i \, h_{\lambda,i}^{*[m]} \right] \\
&= \sum_{i=1}^n  \, h_{\lambda,i}^{*[m]} \, k^{[m-1]}_i
- \sum_{i=1}^n  \,  h_{\lambda,i}^{*[m]} \, k^{[m-1]}_i \, y_i^{k^{[m-1]}_i} \, 
\exp\left(\eta_{\lambda,i}^{[m-1]} \, + \, \nu_\lambda^{[\text{prev}, j]} \, h_{\lambda,i}^{*[m]}\right)^{-k^{[m-1]}_i}
\left[ 1 + k^{[m-1]}_i \, h_{\lambda,i}^{*[m]} \, \nu_\lambda^{[\text{prev}, j]} \right] \\
&\hspace{1.2em} + \nu_\mu^{*[m]} \sum_{i=1}^n  \, {h_{\lambda,i}^{*[m]}}^2 \, {k^{[m-1]}_i}^2 \, y_i^{k^{[m-1]}_i} \, \exp\left(\eta_{\lambda,i}^{[m-1]} \, + \, \nu_\lambda^{[\text{prev}, j]} \, h_{\lambda,i}^{*[m]}\right)^{-k^{[m-1]}_i} \, .
\end{align*}
For the optimal step length we therefore obtain:
\begin{align*}
\nu_\lambda^{*[m]} \approx 
- \frac{\sum_{i=1}^n  \, h_{\lambda,i}^{*[m]} \, k^{[m-1]}_i
- \sum_{i=1}^n  \,  h_{\lambda,i}^{*[m]} \, k^{[m-1]}_i \, y_i^{k^{[m-1]}_i} 
\exp\left(\eta_{\lambda,i}^{[m-1]} \, + \, \nu_\lambda^{[\text{prev}, j]} \, h_{\lambda,i}^{*[m]}\right)^{-k^{[m-1]}_i} \hspace*{-0.2em}
\left[ 1 + k^{[m-1]}_i \, h_{\lambda,i}^{*[m]} \, \nu_\lambda^{[\text{prev}, j]} \right]}
{\sum_{i=1}^n  \, {h_{\lambda,i}^{*[m]}}^2 \, {k^{[m-1]}_i}^2 \, y_i^{k^{[m-1]}_i} \, \exp\left(\eta_{\lambda,i}^{[m-1]} \, + \, \nu_\lambda^{[\text{prev}, j]} \, h_{\lambda,i}^{*[m]}\right)^{-k^{[m-1]}_i}}
\end{align*}

\newpage

\section{Simulation Results}

\subsection{Gaussian location and scale model}
\label{daub:appendix_simu_gaussian}

\begin{figure}[h!]
\includegraphics[width=\textwidth]{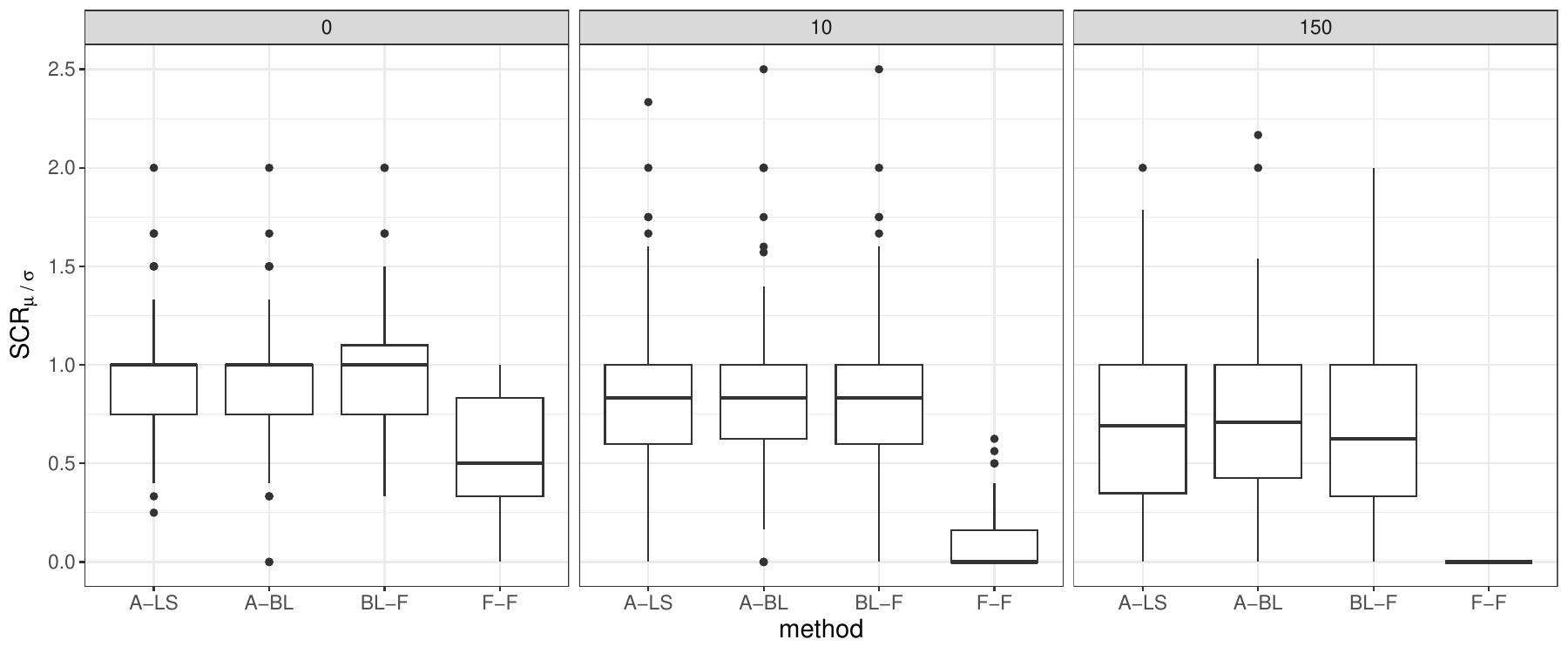}
\caption{\label{daub:fig_overview_selCov_ratio_Gaussian} Distribution of the ratio of numbers of covariates included in $\eta_\mu$ relative to $\eta_\sigma$ (SCR$_{\mu/\sigma}$) in the Gaussian simulation setting (\ref{daub:formula_simulation_model_Gaussian}) for a varying number of additional non-informative covariates (columns).
18 outliers that exceed an SCR$_{\mu/\sigma}$ of 2.5 are regarded in the distributions but not included in the plot for a better comprehensibility.
The excluded outliers occured in 7 simulation runs, mainly with 150 additional non-informative covariates.}
\end{figure}

\begin{figure}[h!]
\includegraphics[width=\textwidth]{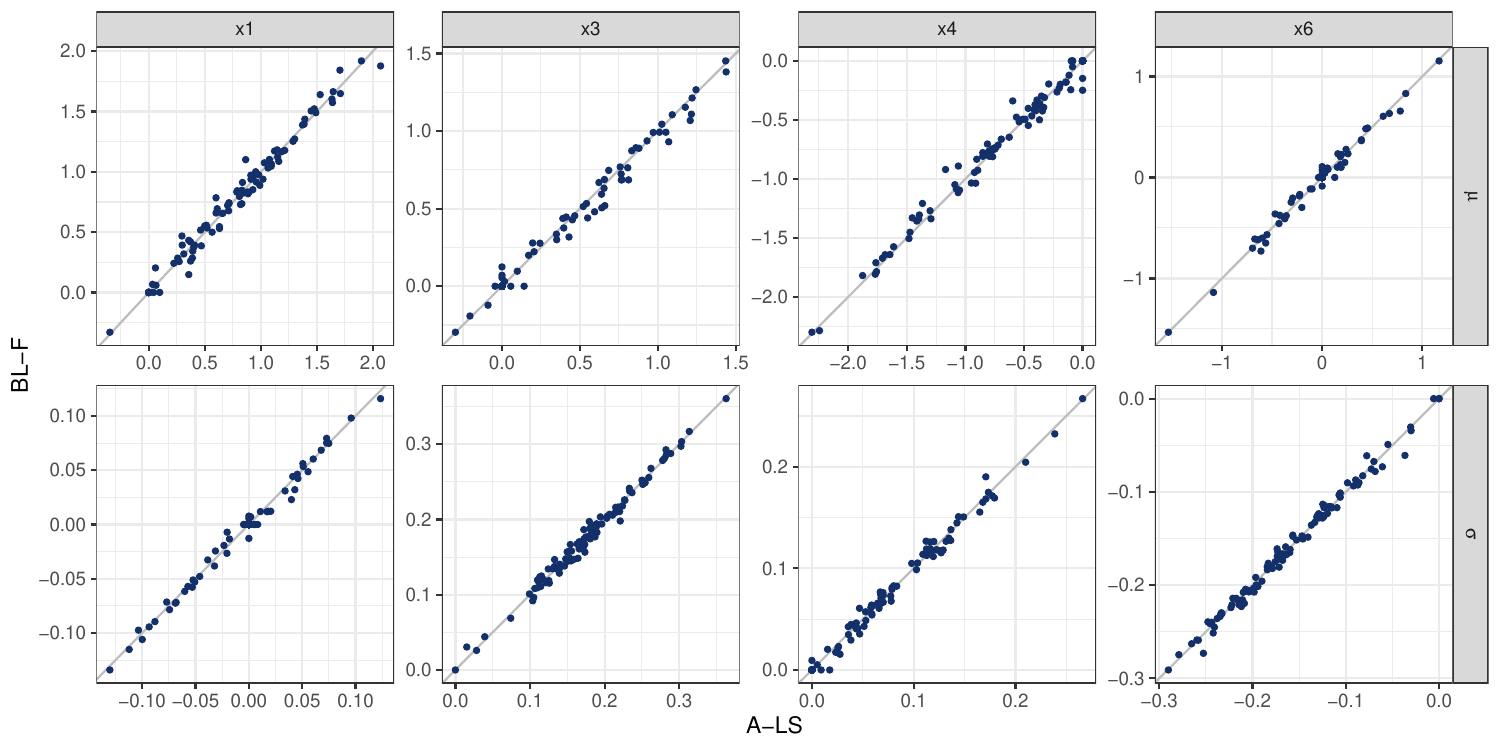}
\caption{\label{daub:fig_scatter_selCoefs_BLF_Gaussian} Direct comparison of coefficient estimates of the two adaptive step length approaches A-LS and BL-F in the Gaussian simulation setting (\ref{daub:formula_simulation_model_Gaussian}) without additional non-informative covariates.}
\end{figure}

\begin{figure}[h!]
\centering
\includegraphics[width=0.8\textwidth]{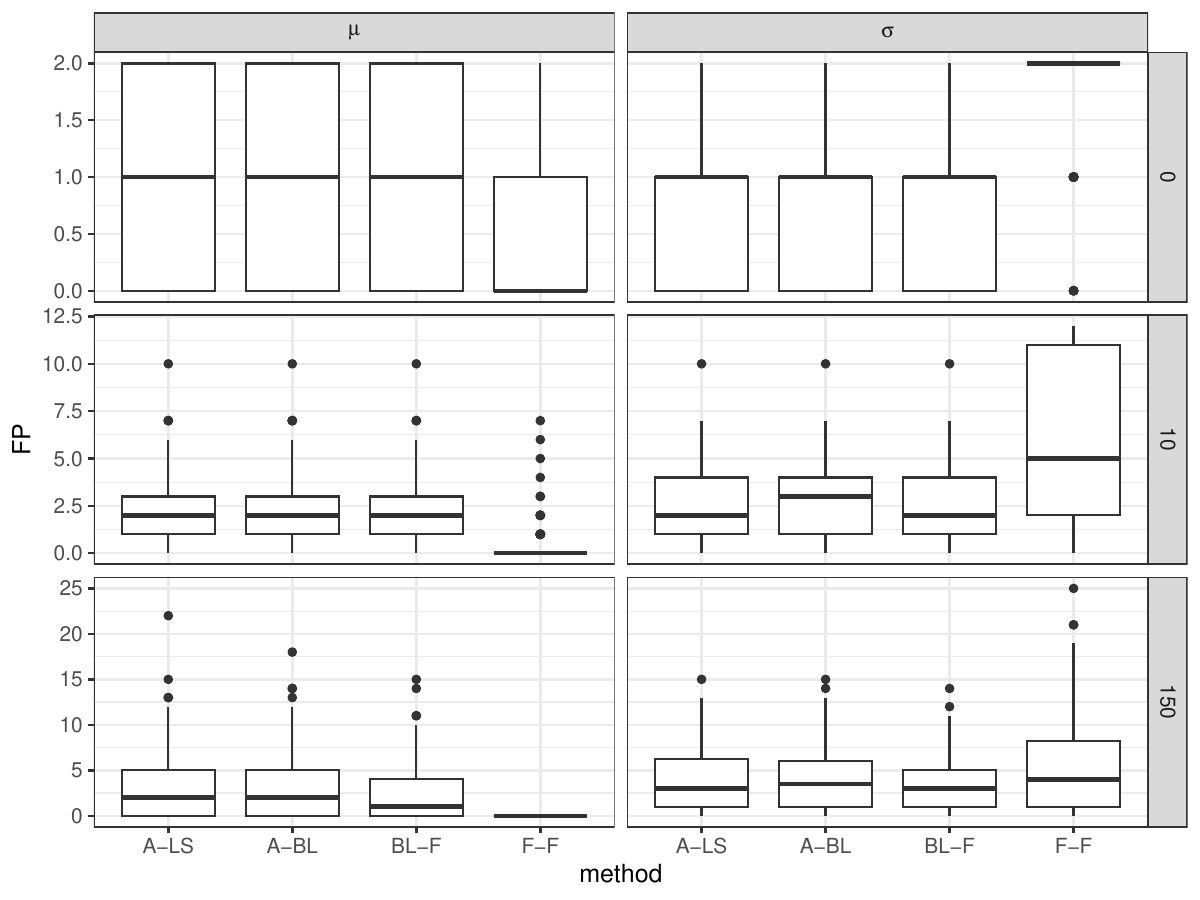}
\caption{\label{daub:fig_overview_FP_Gaussian}
Distribution of the number of false positives (FP) in the Gaussian simulation setting (\ref{daub:formula_simulation_model_Gaussian}) for a varying number of additional non-informative covariates (rows).}
\end{figure}

\begin{figure}[h!]
\centering
\includegraphics[width=0.8\textwidth]{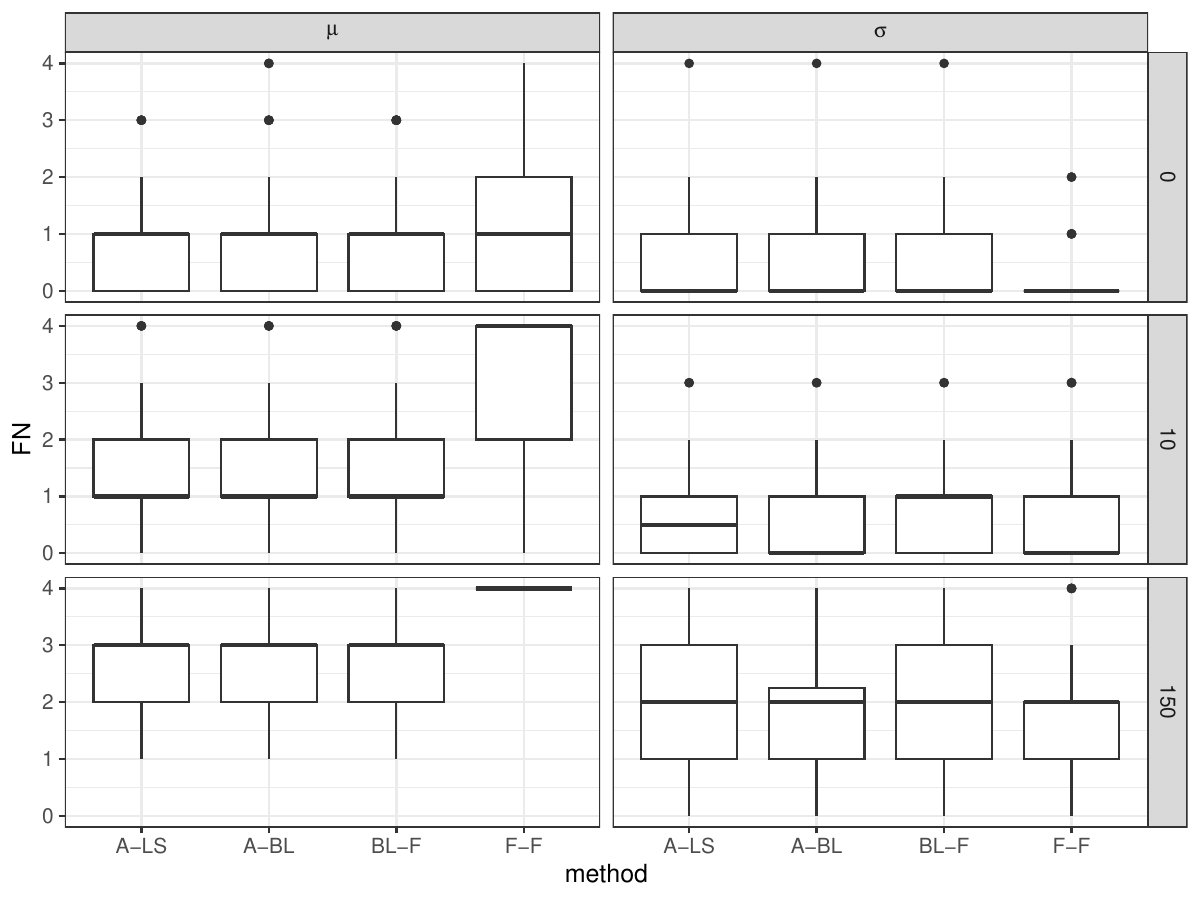}
\caption{\label{daub:fig_overview_FN_Gaussian} Distribution of the number of false negatives (FN) in the Gaussian simulation setting (\ref{daub:formula_simulation_model_Gaussian}) for a varying number of additional non-informative covariates (rows).}
\end{figure}

\clearpage

\newpage

\subsection{Negative binomial location and scale model}
\label{daub:appendix_simu_negbinom}

\begin{figure}[h!]
\centering
\includegraphics[width=0.65\textwidth]{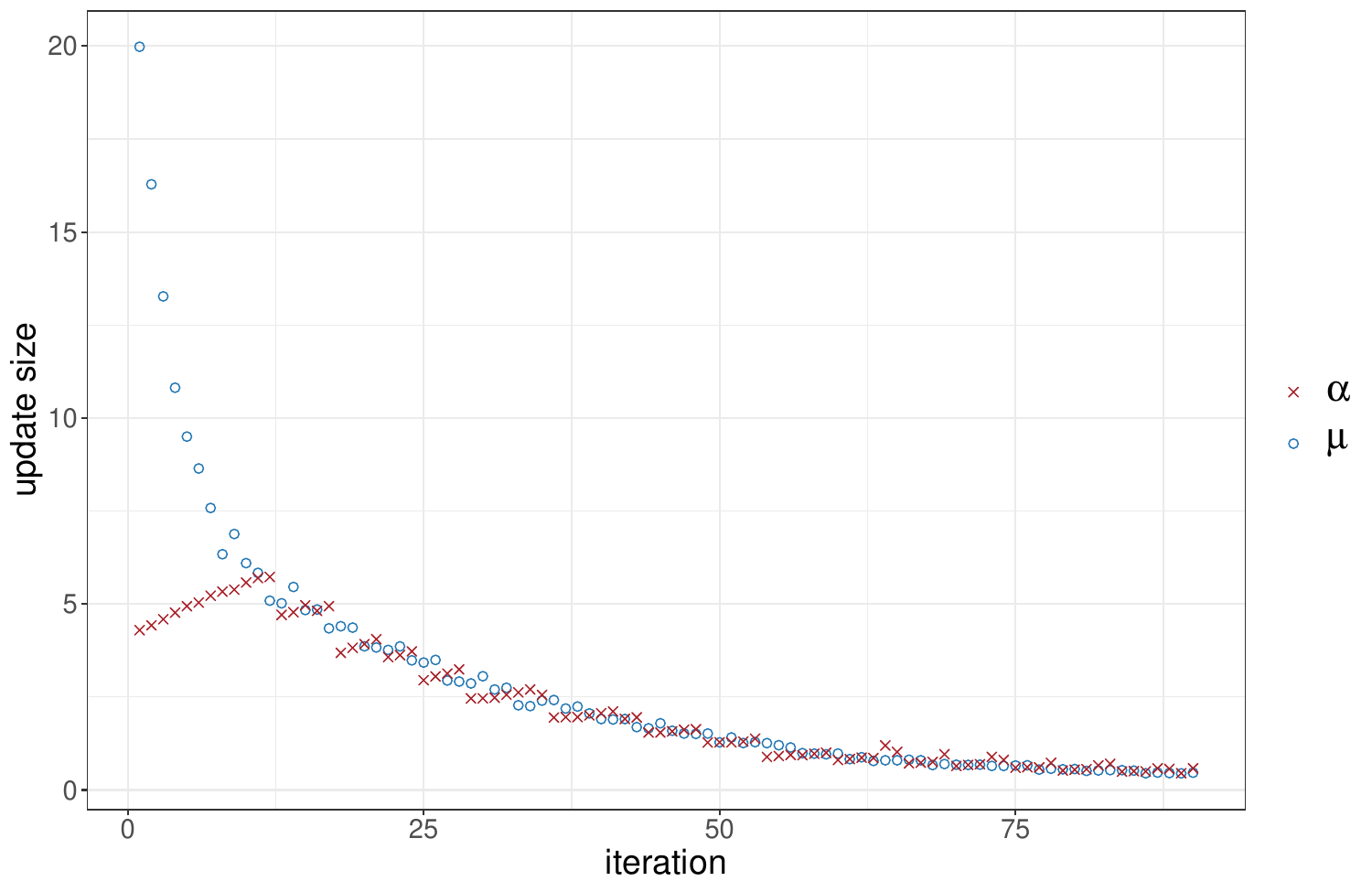}
\caption{\label{daub:fig_stepsizes_single_run_negbinom} Update sizes corresponding to numerically obtained shrunk optimal step lengths for an exemplary simulation run in the negative binomial simulation setting (\ref{daub:formula_simulation_model_negbinom}) without additional non-informative covariates.}
\end{figure}

\begin{figure}[h!]
\centering
\includegraphics[width=0.9\textwidth]{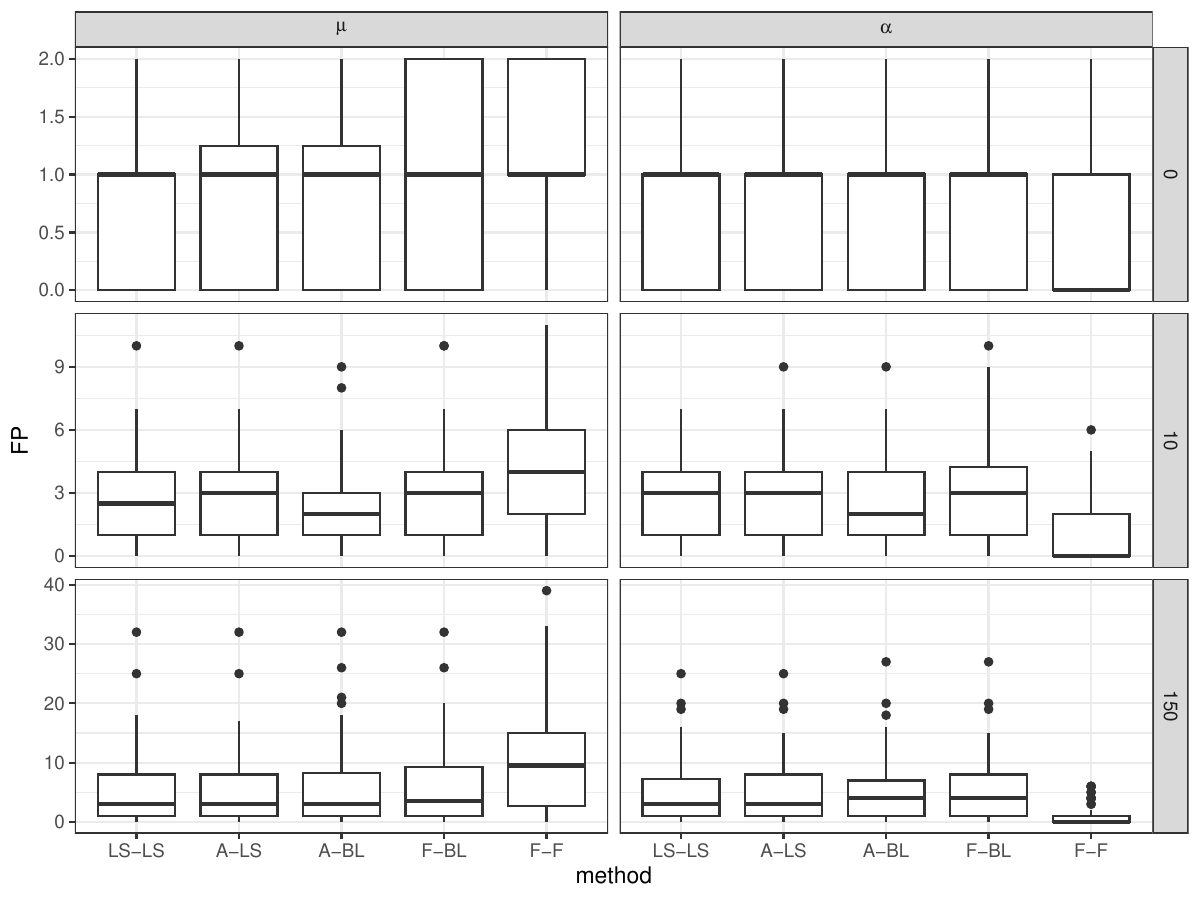}
\caption{\label{daub:fig_overview_FP_negbinom} Distribution of the number of false positives (FP) in the negative binomial simulation setting (\ref{daub:formula_simulation_model_negbinom}) for a varying number of additional non-informative covariates (rows).}
\end{figure}

\begin{figure}[h!]
\centering
\includegraphics[width=0.9\textwidth]{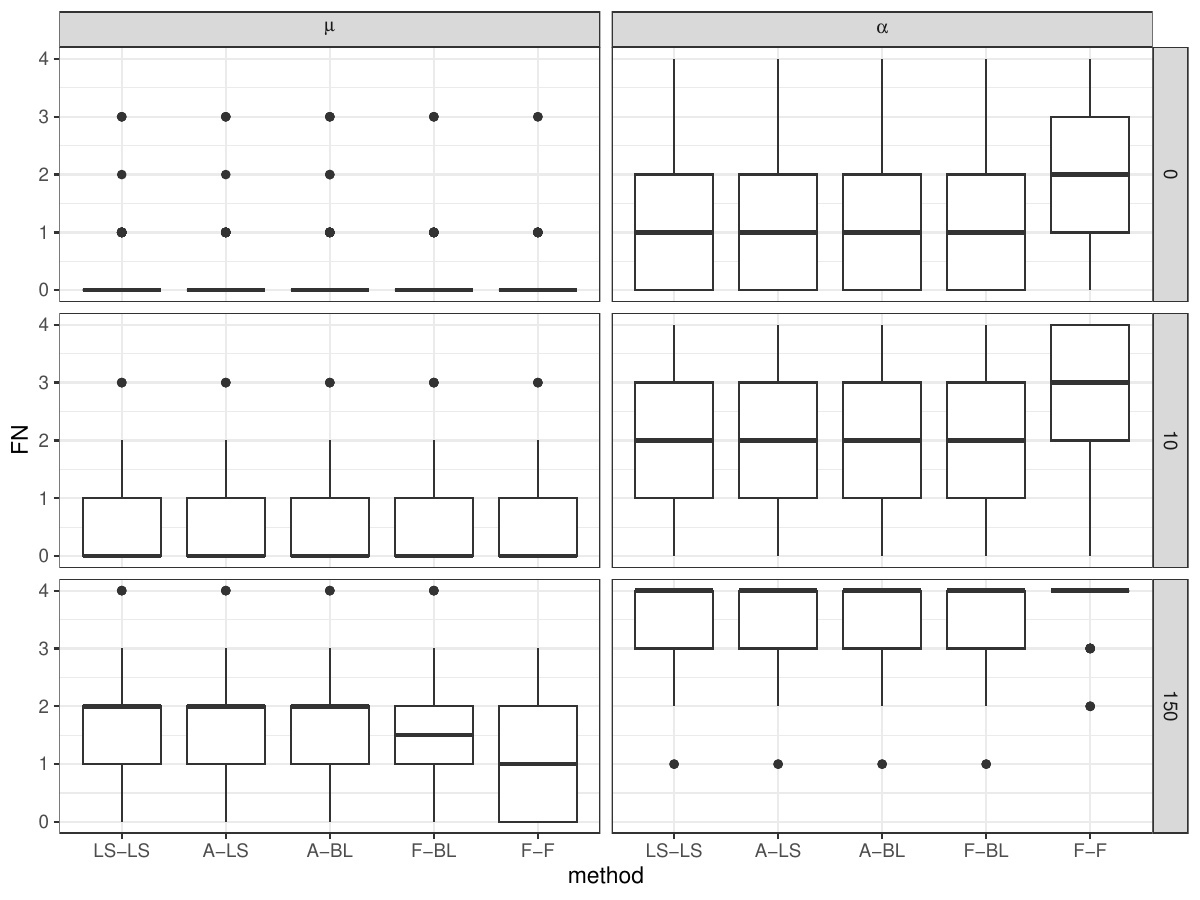}
\caption{\label{daub:fig_overview_FN_negbinom} Distribution of the number of false negatives (FN) in the negative binomial simulation setting (\ref{daub:formula_simulation_model_negbinom}) for a varying number of additional non-informative covariates (rows).}
\end{figure}

\begin{figure}[h!]
\includegraphics[width=\textwidth]{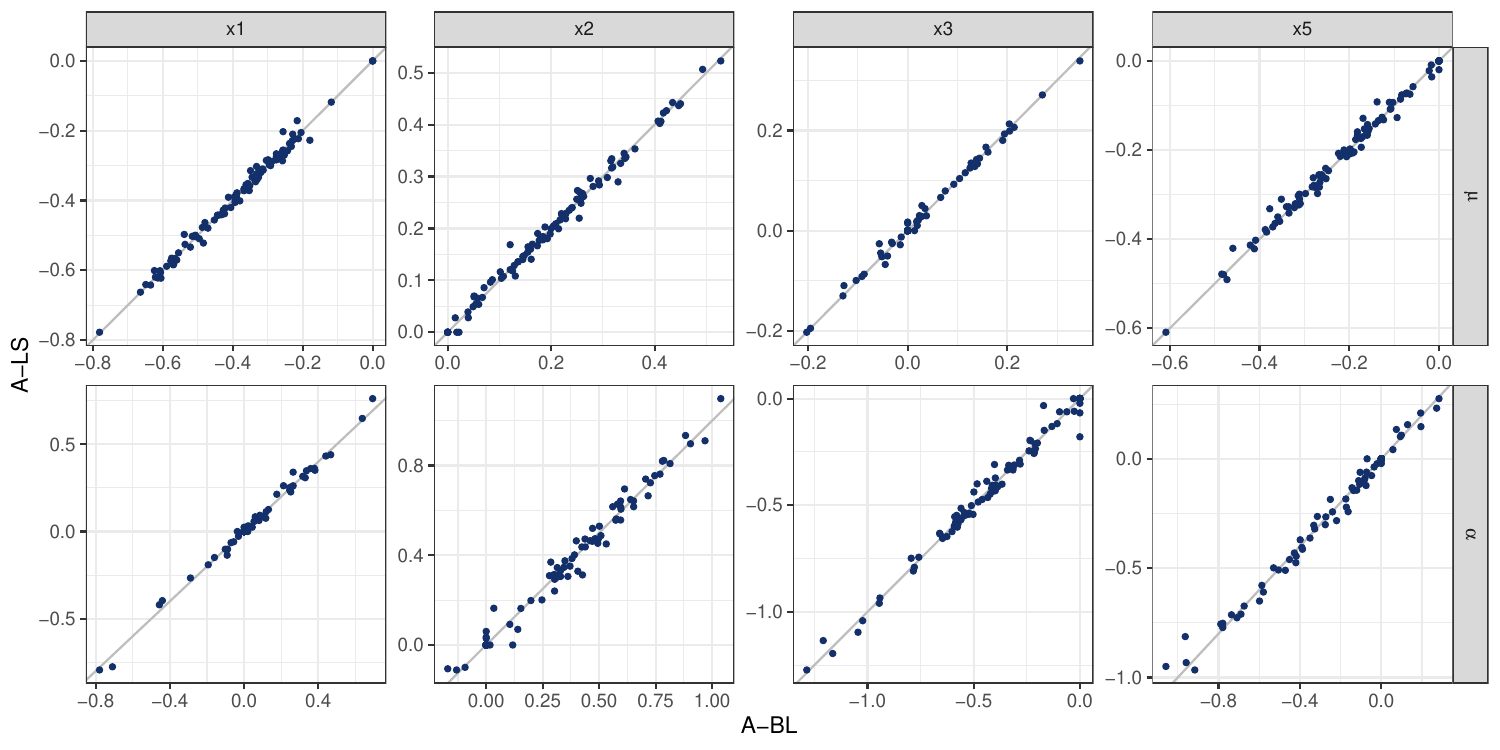}
\caption{\label{daub:fig_scatter_selCoefs_ALAD_ALBL_negbinom} Direct comparison of coefficient estimates of the two adaptive step length approaches A-LS and A-BL in the negative binomial simulation setting (\ref{daub:formula_simulation_model_negbinom}) without additional non-informative covariates.}
\end{figure}

\clearpage

\begin{figure}[h!]
\includegraphics[width=\textwidth]{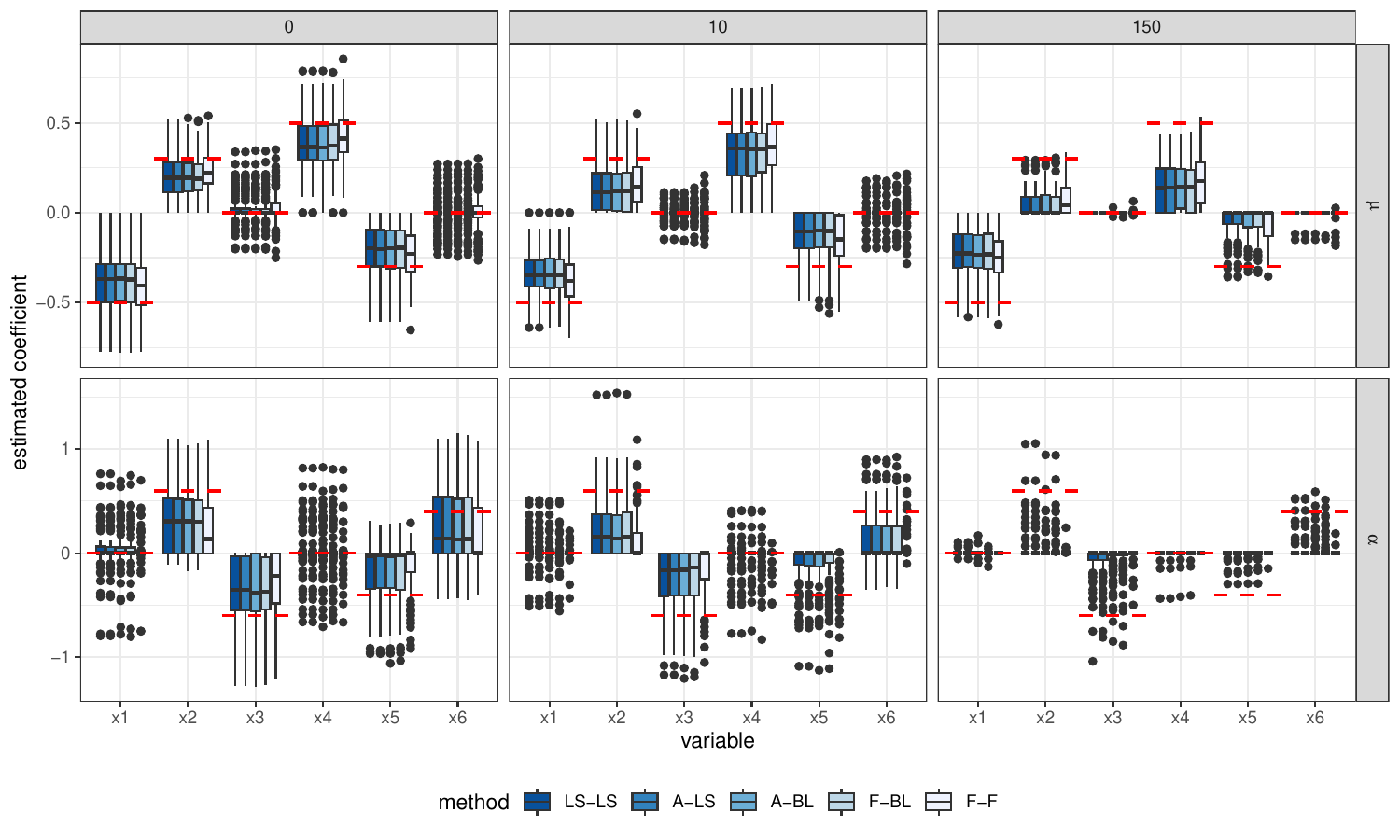}
\caption{\label{daub:fig_overview_coeffs_negbinom} Distribution of the coefficient estimates in the negative binomial simulation setting (\ref{daub:formula_simulation_model_negbinom}) for a varying number of additional non-informative covariates (columns).
The red dashed lines represent the true coefficients.}
\end{figure}

\clearpage

\subsection{Weibull scale and shape model} 
\label{daub:appendix_simu_weibull}

We consider the following model for the Weibull response variable $y_i \sim \mathcal{WB}(\lambda_i, k_i)$
\begin{align}
\eta_{\lambda, i} &= \log(\lambda_i) = 0.6 + 0.15 \cdot x_{1i} - 0.2 \cdot x_{2i} + 0.4 \cdot x_{3i} - 0.25 \cdot x_{4i} \nonumber \\
\eta_{k, i} &= \log(k_i) = - 0.15 \cdot x_{3i} + 0.15 \cdot x_{4i} - 0.1 \cdot x_{5i} + 0.1 \cdot x_{6i} ,
\label{daub:formula_simulation_model_weibull}
\end{align}
where $x_1,...,x_6$ are drawn independently from a uniform distribution on $[-1,1]$. 
In this simulation setting, 0, 10, 150 additional non-informative covariates are considered. \\

\begin{figure}[h!]
\includegraphics[width=\textwidth]{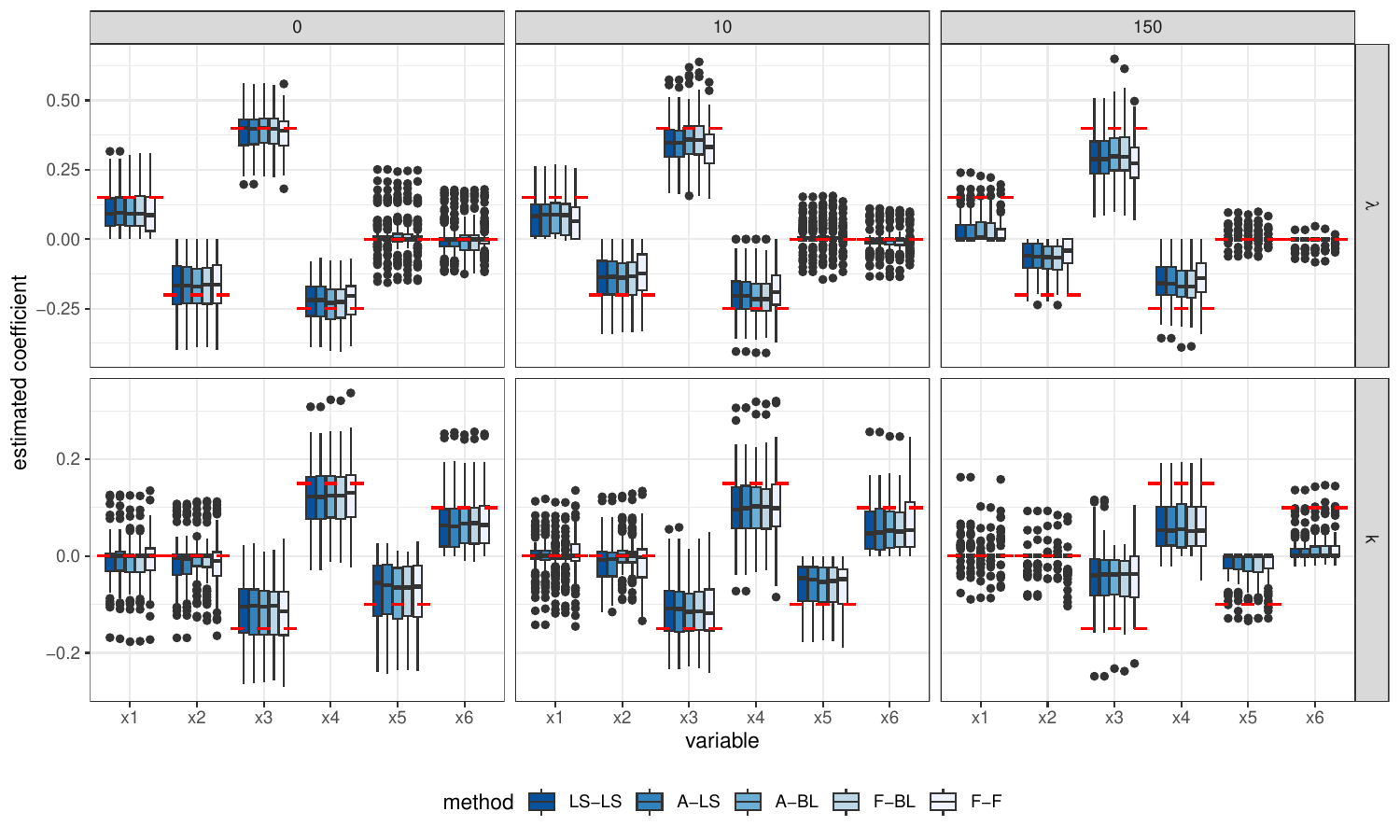}
\caption{\label{daub:fig_overview_coeffs_weibull} Distribution of the coefficient estimates in the Weibull simulation setting (\ref{daub:formula_simulation_model_weibull}) for a varying number of additional non-informative covariates (columns).
The red dashed lines represent the true coefficients.}
\end{figure}

\begin{figure}[!htb] 
\includegraphics[width=\textwidth]{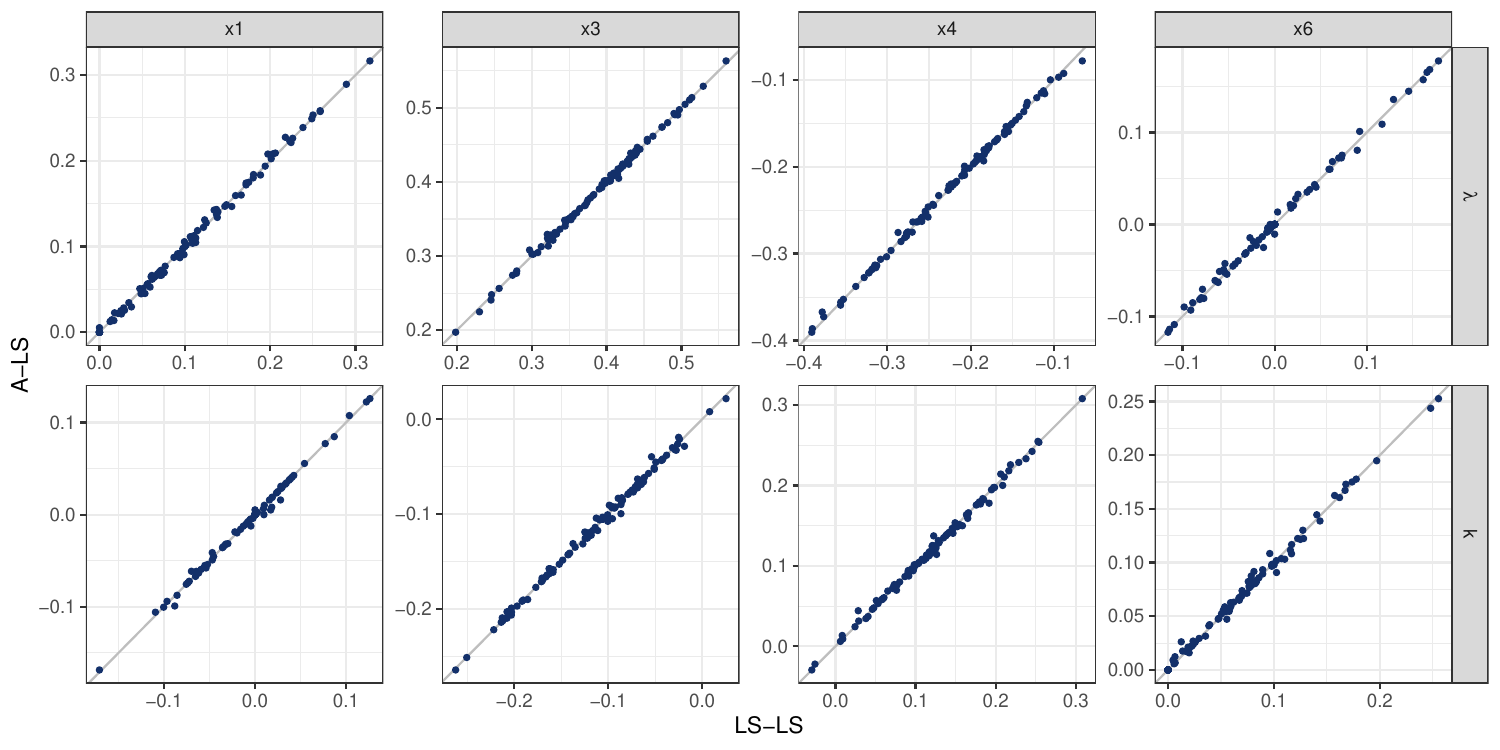}
\caption{\label{daub:fig_scatter_selCoefs_weibull} Direct comparison of coefficient estimates of the two adaptive step length approaches LS-LS and A-LS in the Weibull simulation setting (\ref{daub:formula_simulation_model_weibull}) without additional non-informative covariates.}
\end{figure}

\begin{figure}[!htb] 
\includegraphics[width=\textwidth]{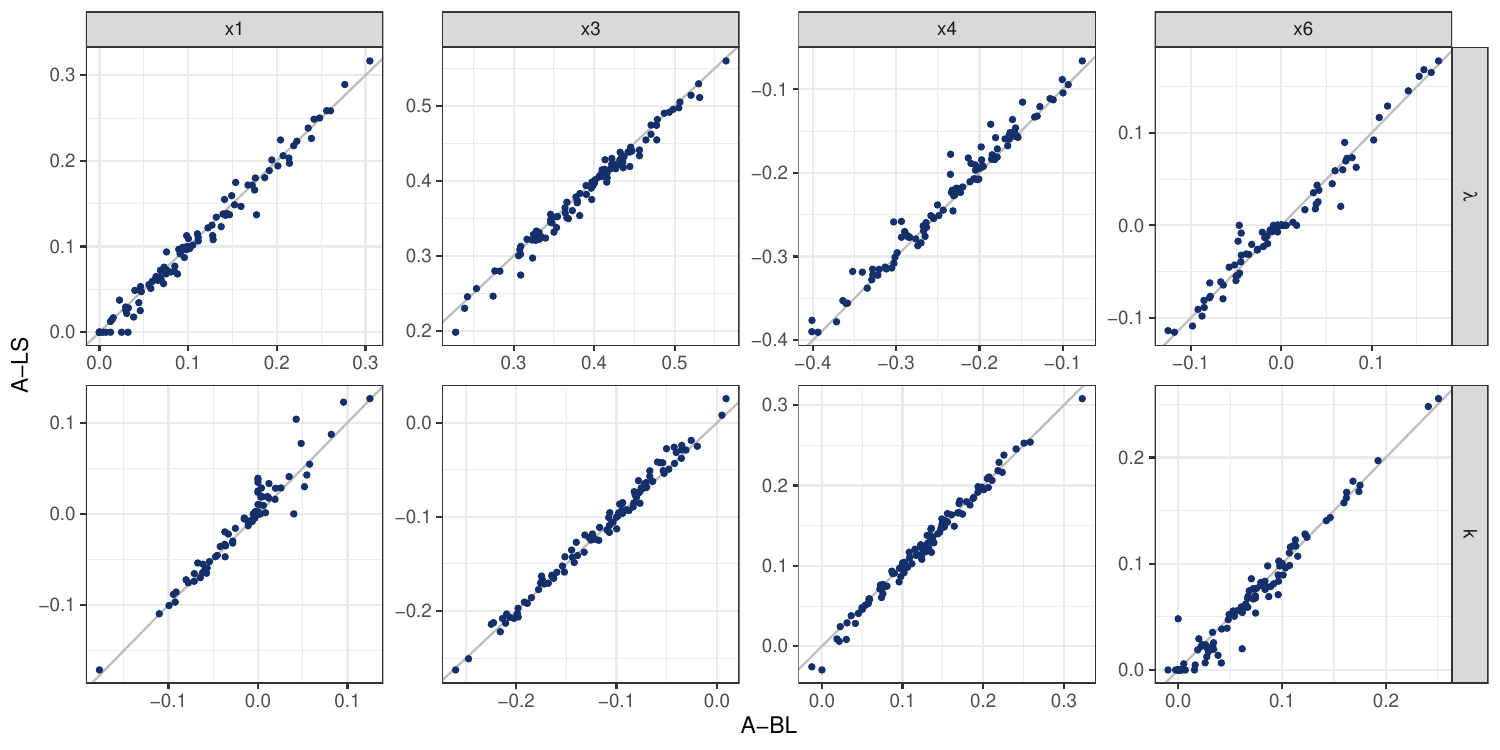}
\caption{\label{daub:fig_scatter_selCoefs_weibull_ALS_ABL} Direct comparison of coefficient estimates of the two adaptive step length approaches A-LS and A-BL in the Weibull simulation setting (\ref{daub:formula_simulation_model_weibull}) without additional non-informative covariates.}
\end{figure}

\begin{figure}[!htb] 
\includegraphics[width=\textwidth]{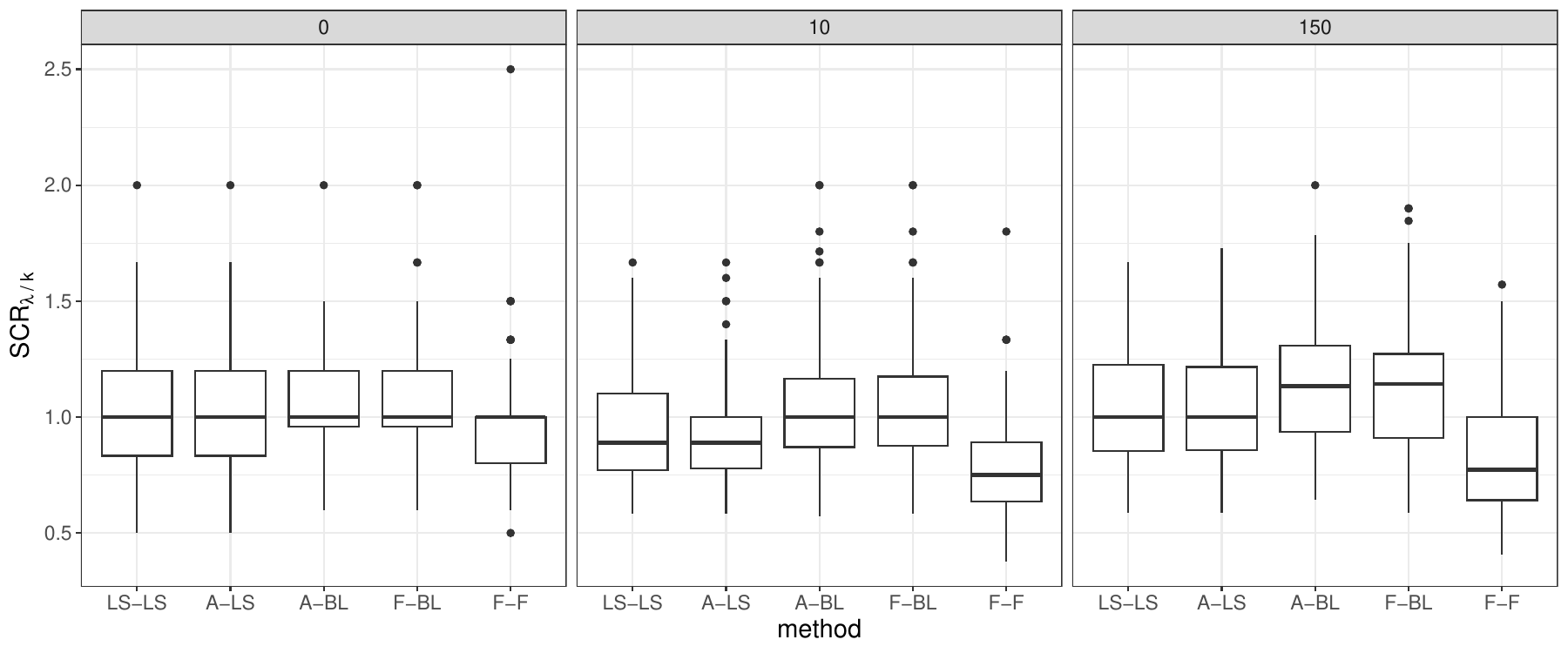}
\caption{\label{daub:fig_overview_selCoefs_raio_weibull} Distribution of the ratio of numbers of covariates included in $\eta_\lambda$ relative to $\eta_k$ (SCR$_{\lambda/k}$) in the Weibull simulation setting (\ref{daub:formula_simulation_model_weibull}) for a varying number of additional non-informative covariates (columns).}
\end{figure}

\begin{table}[h!] 
      \centering
\begin{tabular}{llllll}
  \toprule[0.09 em]
 & \hspace*{0.4em} LS-LS & \hspace*{0.6em} A-LS & \hspace*{0.6em} A-BL & \hspace*{0.6em} F-BL & \hspace*{1.2em} F-F \\ 
  \midrule
1st Qu. & 0.394 (106) & 0.340 (108) & 0.346 (140) & 0.199 (101) & 0.179 (90) \\ 
  Median & 0.579 (128) & 0.438 (130) & 0.450 (159) & 0.277 (120) & 0.256 (108) \\ 
  3rd Qu. & 0.757 (181) & 0.537 (180) & 0.581 (196) & 0.393 (151) & 0.386 (146) \\   
   \bottomrule[0.09 em]
\end{tabular}
\caption{\label{daub:table_run_time_weibull}
Quartiles of the run times until $m_\text{stop}$ in seconds and stopping iterations (in parenthesis) in the Weibull binomial simulation setting (\ref{daub:formula_simulation_model_weibull}) without additional non-informative covariates.}
\end{table}

\clearpage

\newpage

\section{Application Results}

\subsection{Australian Health Care Data}
\label{daub:appendix_healthAustralia}

\subsubsection{Information on the data}
\label{daub:appendix_info_data_healthAustralia}

\begin{table}[h!]
\centering
\begin{tabular}{ll}
  \textbf{doctorco} & Number of consultations with a doctor or specialist in the past 2 weeks. \\[0.3em]
  \textbf{sex} & 1 if female, 0 if male. \\[0.3em]
  \textbf{age} & Age in years divided by 100. \\[0.3em]
  \textbf{income} & Annual income in Australian dollars divided by 1000. \\[0.3em]
  \textbf{illness} & Number of illnesses in the past 2 weeks, 0 to 5 or more. \\[0.3em] 
  \textbf{actdays} & Number of days of reduced activity in the past two weeks due to illness or injury. \\[0.3em]
  \textbf{hospadmi} & Number of admissions to a hospital, psychiatric hospital, nursing or convalescent home in \\
  & the past 12 months (up to 5 or more admissions). \\[0.3em]
  \textbf{hospdays} & Number of nights in a hospital, etc. during most recent admission, 
  0 if no admission in the \\
  & past 12 months. \\[0.3em]
  \textbf{medicine} & Total number of prescribed and non-prescribed medications used in past 2 days. \\[0.3em]
  \textbf{prescrib} & Total number of prescribed medications used in past 2 days. \\[0.3em]
  \textbf{nondocco} & Number of consultations with non-doctor health professionals in the past 4 weeks. \\[0.3em] 
  \textbf{levyplus} & 1 if covered by private health insurance fund for private patient in public hospital (with doc- \\
  & tor of choice), 0 otherwise. \\[0.3em]
  \textbf{freepoor} & 1 if covered free by government because low income, recent immigrant, unemployed, 0 other-\\
  & wise. \\[0.3em]
  \textbf{freepera} & 1 if covered free by government because of old age, disability pension, invalid veteran or fa- \\
  & mily of deceased veteran, 0 otherwise. \\[0.3em]   
   \textbf{hscore} & General health questionnaire score using Goldberg's method. High score indicates bad health. \\[0.3em]
  \textbf{chcond1} & 1 if chronic condition(s) but not limited in activity, 0 otherwise. \\[0.3em]
  \textbf{chcond2} & 1 if chronic condition(s) and limited in activity, 0 otherwise. 
\end{tabular}
\end{table}

\noindent This overview follows \cite{Cameron1988} and \cite{Karlis2005}. For more a more detailed description of the data, see \cite{Cameron1988}.

\subsubsection{Correlation plot}
\label{daub:appendix_healthAustralia_corr}
\begin{figure}[h!]
\center
\includegraphics[width=0.6\textwidth]{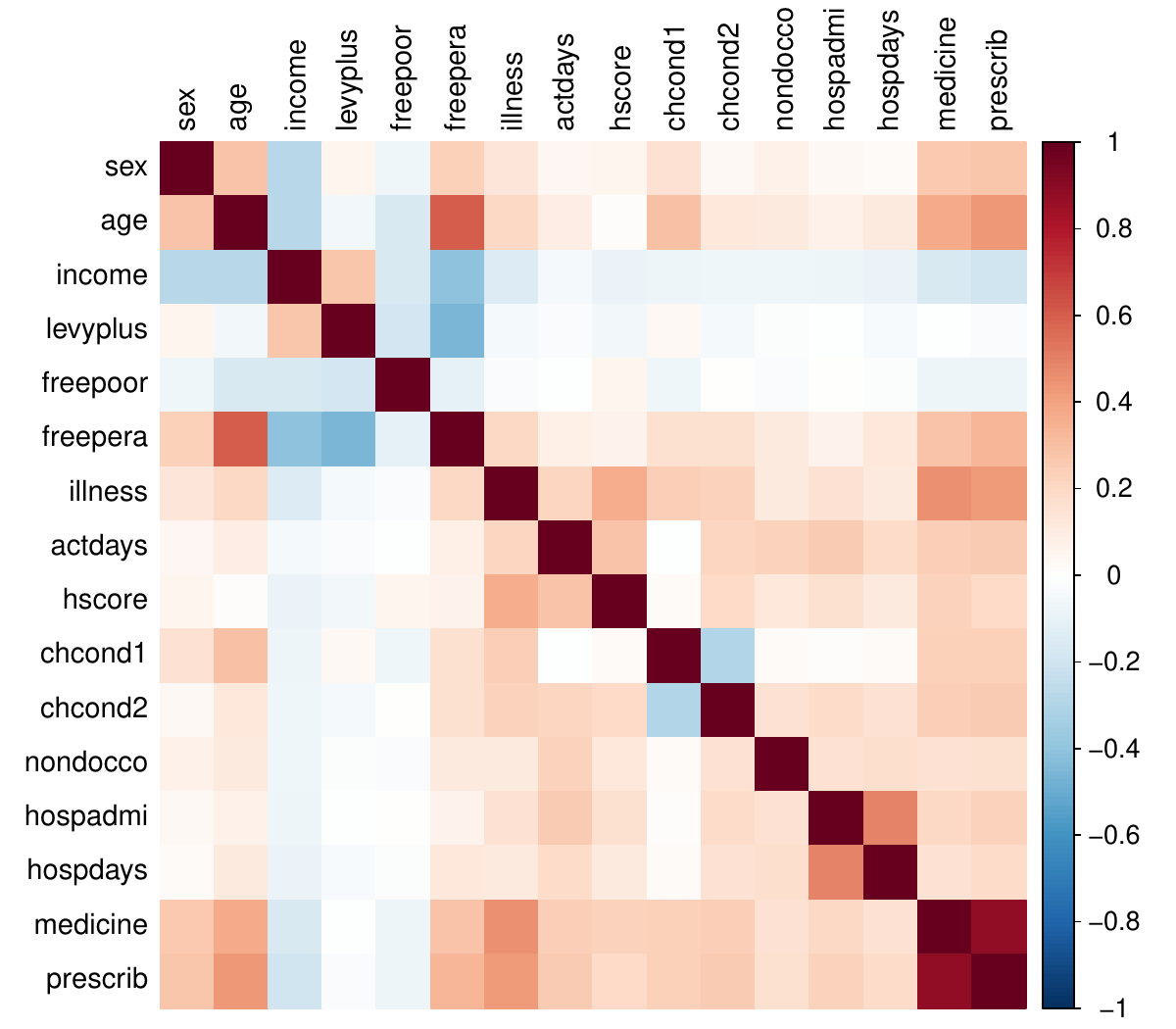}
\caption{\label{daub:fig_healthAustralia_corr_plot} Correlation plot of the Australia health care data.} 
\end{figure}

\newpage

\subsubsection{Technical properties of boosting runs}
\label{daub:appendix_healthAustralia_technical_properties}

\begin{table}[h!] 
      \centering
      \begin{flushleft}
      \end{flushleft}
      \vspace*{-1em}
\begin{tabular}{lllll}
  \toprule[0.09 em]
 & \hspace*{0.6em} A-LS & \hspace*{0.6em} A-BL & \hspace*{1.2em} F-F \\ 
  \midrule
1st Qu. & 2.92 (83) & 2.46 (120) & 18.31 (975) \\ 
  Median & 3.28 (103) & 2.71 (142) & 18.66 (1,084) \\ 
  3rd Qu. & 5.09 (151) & 4.17 (220) & 23.48 (1,241) \\ 
   \bottomrule[0.09 em]
\end{tabular}
\caption{\label{daub:table_run_time_stopping_it_HealthAustralia}
Quartiles of the run times until $m_\text{stop}$ in seconds and stopping iterations (in parenthesis) for 100 cross-validation runs for the Australian health care data.}
\end{table}

\subsubsection{Results for LS-LS and F-BL step lengths}
\label{daub:appendix_Health_Australia_LS-LS_F-BL}

\begin{figure}[!htb] 
\includegraphics[width=\textwidth]{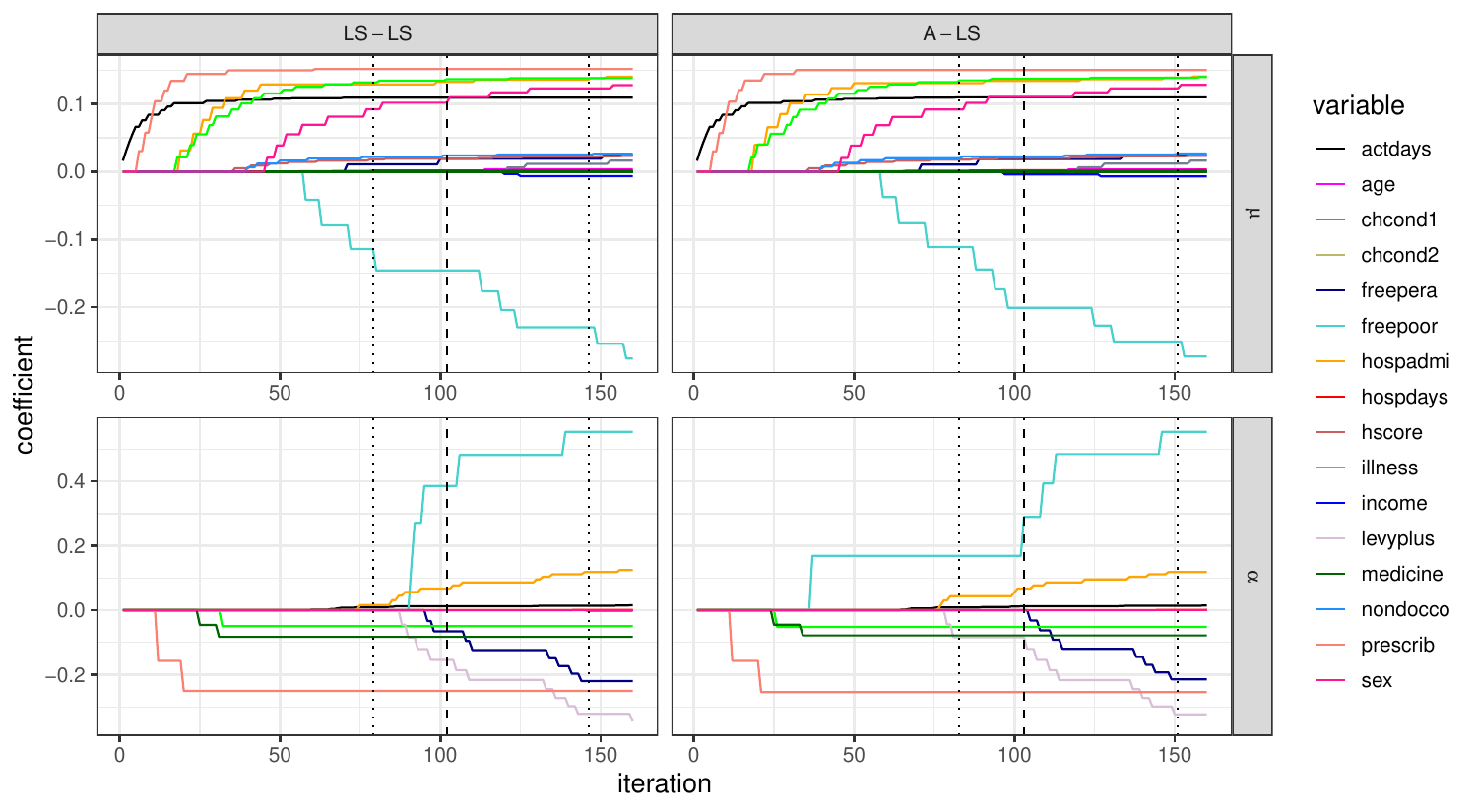}
\caption{\label{daub:fig_healthAustralia_coeff_paths_LSLS_ALS} Comparison of the coefficient paths for the Australian health care data using LS-LS and A-LS step lengths (columns). 
The vertical lines represent the median stopping iteration (dashed) as well as the 1st and 3rd quartile (dotted) of the stopping iteration obtained via cross-validation on 100 randomly drawn folds.} 
\end{figure}

\begin{figure}[!htb] 
\includegraphics[width=\textwidth]{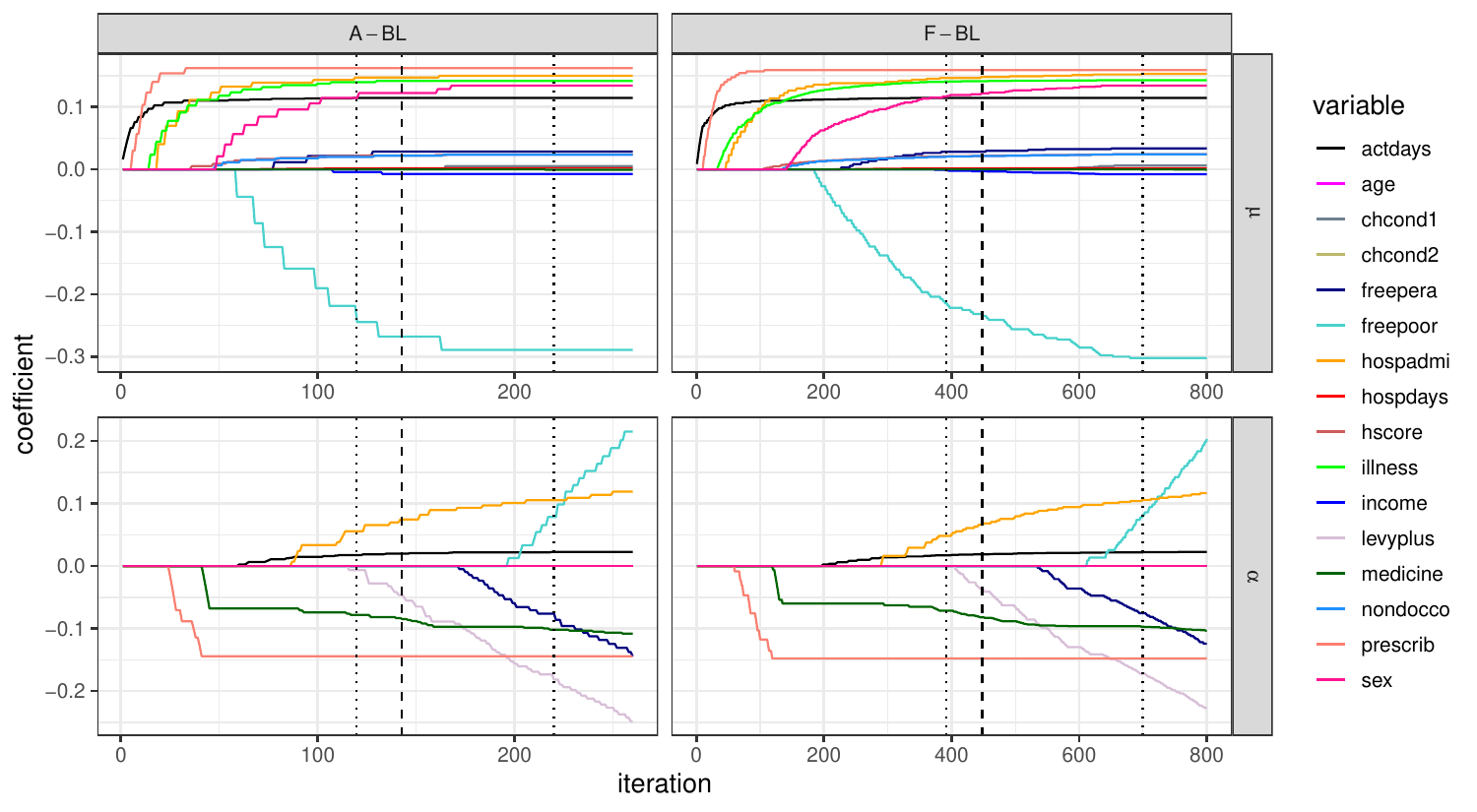}
\caption{\label{daub:fig_healthAustralia_coeff_paths_ABL_FBL} Comparison of the coefficient paths for the Australian health care data using A-BL and F-BL step lengths (columns). 
The vertical lines represent the median stopping iteration (dashed) as well as the 1st and 3rd quartile (dotted) of the stopping iteration obtained via cross-validation on 100 randomly drawn folds.
Please be aware of the differences in the x-axis scaling.} 
\end{figure}

\begin{table}[!htb] 
\centering
\begin{tabular}{c|ccccc|ccccc}
  \toprule[0.09 em]
  & & & $\eta_\mu$ & & & & & $\eta_\alpha$ & & \\[0.3em]
 & LS-LS & A-LS & A-BL & F-BL & F-F & LS-LS & A-LS & A-BL & F-BL & F-F \\ 
  \midrule
(Intercept) & -1.998 & -2.009 & -2.062 & -2.054 & -2.099 & 0.729 & 0.696 & 0.346 & 0.347 & 0.257 \\ 
  sex & 0.102 & 0.110 & 0.122 & 0.122 & 0.142 & 0 & 0 & 0 & 0 & 0 \\ 
  age & 0 & 0 & 0 & 0 & 0 & 0 & 0 & 0 & 0 & 0 \\ 
  income & 0 & -0.004 & -0.007 & -0.003 & -0.012 & 0 & 0 & 0 & 0 & 0 \\ 
  illness & 0.137 & 0.137 & 0.142 & 0.141 & 0.148 & -0.049 & -0.052 & 0 & 0 & 0 \\ 
  actdays & 0.109 & 0.110 & 0.114 & 0.114 & 0.117 & 0.012 & 0.012 & 0.020 & 0.019 & 0.022 \\  
  hospadmi & 0.132 & 0.134 & 0.147 & 0.148 & 0.158 & 0.067 & 0.068 & 0.075 & 0.066 & 0.063 \\ 
  hospdays & 0.002 & 0.002 & 0.002 & 0.002 & 0.002 & 0 & 0 & 0 & 0 & 0 \\ 
  medicine & 0 & 0 & 0 & 0 & -0.016 & -0.083 & -0.078 & -0.083 & -0.082 & -0.100 \\ 
  prescrib & 0.152 & 0.150 & 0.162 & 0.159 & 0.177 & -0.250 & -0.254 & -0.144 & -0.148 & -0.089 \\  
  nondocco & 0.024 & 0.022 & 0.022 & 0.021 & 0.023 & 0 & 0 & 0 & 0 & 0 \\ 
  levyplus & 0 & 0 & 0 & 0 & 0 & -0.154 & -0.084 & -0.046 & -0.036 & 0 \\ 
  freepoor & -0.146 & -0.201 & -0.268 & -0.235 & -0.347 & 0.386 & 0.290 & 0 & 0 & 0 \\ 
  freepera & 0.019 & 0.019 & 0.028 & 0.028 & 0.034 & -0.065 & 0 & 0 & 0 & 0 \\  
  hscore & 0.019 & 0.020 & 0.022 & 0.022 & 0.025 & 0 & 0 & 0 & 0 & 0 \\ 
  chcond1 & 0 & 0 & 0 & 0 & 0.008 & 0 & 0 & 0 & 0 & 0 \\ 
  chcond2 & 0 & 0 & 0 & 0 & 0 & 0 & 0 & 0 & 0 & 0 \\
  \bottomrule[0.09 em]
\end{tabular}
\caption{\label{daub:table_coefficient_estimates_all} Coefficient estimates at the median stopping iteration for the Australian health care data applying different step length approaches.}
\end{table}

\clearpage

\subsection{Diffuse Large-B-cell Lymphoma Data}
\label{daub:appendix_DLBCL}

\begin{table}[htb]
\centering
\begin{tabular}{c|ccccc}
  \toprule[0.09 em]
  & LS-LS & A-LS & A-BL & F-BL & F-F \\[0.3em]
  \midrule
  (Intercept) & 1.056 & 1.054 & 1.054 & 1.047 & 0.857 \\
  X28641 & 0.061 & 0.061 & 0.073 & 0.052 & 0.088 \\  
  X24683 & 0.031 & 0.031 & 0.057 & 0.074 & 0.019 \\ 
  X16799 & 0.043 & 0.044 & 0.040 & 0.050 & 0.064 \\ 
  X30240 & -0.040 & -0.042 & -0.167 & -0.224 & 0 \\ 
  X16545 & 0.036 & 0.037 & 0.066 & 0.063 & 0 \\ 
  X34688 & -0.109 & -0.110 & -0.106 & 0 & -0.099 \\ 
  X28673 & -0.087 & -0.087 & 0 & 0 & -0.102 \\ 
  X30109 & 0.019 & 0.019 & 0 & 0 & 0.021 \\ 
  X26321 & -0.072 & -0.072 & 0 & 0 & -0.071 \\ 
  X28910 & -0.037 & -0.037 & 0 & 0 & 0 \\ 
  X32118 & 0 & 0 & -0.043 & -0.083 & 0 \\ 
  X15937 & 0 & 0 & -0.109 & -0.143 & 0 \\ 
  X27593 & 0 & 0 & 0.020 & 0 & 0.020 \\ 
  X33186 & 0 & 0 & -0.043 & 0 & 0 \\ 
  X16264 & 0 & 0 & 0 & 0.048 & 0 \\ 
  X24816 & 0 & 0 & 0 & 0 & -0.051 \\ 
  X24412 & 0 & 0 & 0 & 0 & -0.049 \\ 
  X27341 & 0 & 0 & 0 & 0 & 0.048 \\ 
  X33198 & 0 & 0 & 0 & 0 & 0.093 \\     
  \bottomrule[0.09 em]
\end{tabular}
\caption{\label{daub:table_coeffs_lambda_DLBCL} 
Coefficient estimates of $\eta_\lambda$ for the DLBCL data applying different step length approaches (columns).}
\end{table}

\begin{table}[htb]
\centering
\begin{tabular}{c|ccccc}
  \toprule[0.09 em]
  & LS-LS & A-LS & A-BL & F-BL & F-F \\[0.3em]
  \midrule
(Intercept) & 0.306 & 0.306 & 0.509 & 0.493 & 0.091 \\ 
  X34729 & 0.046 & 0.046 & 0.059 & 0.074 & 0.067 \\ 
  X29181 & 0.035 & 0.035 & 0.048 & 0.085 & 0 \\ 
  X26250 & 0.054 & 0.054 & 0.013 & 0 & 0.164 \\
  X27129 & -0.102 & -0.102 & 0 & 0 & -0.140 \\ 
  X17180 & 0.020 & 0.020 & 0 & 0 & 0 \\ 
  X27593 & 0.021 & 0.021 & 0 & 0 & 0 \\  
  X24325 & 0 & 0 & -0.005 & -0.013 & 0 \\ 
  X34552 & 0 & 0 & 0.052 & 0.039 & 0 \\ 
  X26012 & 0 & 0 & -0.027 & 0 & 0 \\  
  X30727 & 0 & 0 & 0.004 & 0 & 0 \\ 
  X28178 & 0 & 0 & 0 & 0 & 0.036 \\ 
  \bottomrule[0.09 em]
\end{tabular}
\caption{\label{daub:table_coeffs_k_DLBCL} Coefficient estimates of $\eta_k$ for the DLBCL data applying different step length approaches (columns).}
\end{table}